\documentclass[aps,reprint,prl,longbibliography,superscriptaddress]{revtex4-1}
\usepackage{}
\usepackage{amssymb}
\usepackage{amsmath}
\usepackage{dcolumn}
\usepackage{bm}
\usepackage{graphicx}
\usepackage{mathrsfs}
\usepackage[colorlinks,linkcolor=blue,anchorcolor=blue,citecolor=blue,urlcolor=blue]{hyperref}

\begin{document}

\title{Enhancing spin-phonon and spin-spin interactions using linear resources in a hybrid quantum system}
\author{Peng-Bo Li}
\affiliation{MOE Key Laboratory for Nonequilibrium Synthesis and Modulation of Condensed Matter, Shaanxi Province Key Laboratory of Quantum Information and Quantum Optoelectronic Devices, School  of  Physics, Xi'an Jiaotong University, Xi'an 710049, China }
\affiliation {Theoretical Quantum Physics Laboratory, RIKEN Cluster for Pioneering Research, Wako-shi, Saitama 351-0198, Japan}
\author{Yuan Zhou}
\affiliation{MOE Key Laboratory for Nonequilibrium Synthesis and Modulation of Condensed Matter, Shaanxi Province Key Laboratory of Quantum Information and Quantum Optoelectronic Devices, School  of  Physics, Xi'an Jiaotong University, Xi'an 710049, China }
\affiliation{School of Science, Hubei University of Automotive Technology, Shiyan 442002, China}
\affiliation{ Division of Physics and Applied Physics, School of Physical and Mathematical Sciences, Nanyang Technological University, Singapore 637371, Singapore}
\author{Wei-Bo Gao}
 \affiliation{ Division of Physics and Applied Physics, School of Physical and Mathematical Sciences,
Nanyang Technological University, Singapore 637371, Singapore}
\author{Franco Nori}
\affiliation {Theoretical Quantum Physics Laboratory, RIKEN Cluster for Pioneering Research, Wako-shi, Saitama 351-0198, Japan}
\affiliation {Department of Physics, The University of Michigan, Ann Arbor, Michigan 48109-1040, USA}

\date{\today}

\begin{abstract}
Hybrid spin-mechanical setups offer a versatile platform for quantum science and technology, but  improving the spin-phonon as well as the spin-spin couplings of such systems remains a crucial challenge. Here, we propose and analyze an experimentally feasible and simple method for exponentially enhancing the spin-phonon, and the phonon-mediated spin-spin interactions in a hybrid spin-mechanical setup, using only \emph{ linear resources}.
Through modulating the spring constant of the mechanical cantilever  with a time-dependent pump, we can acquire a tunable and  nonlinear (two-phonon) drive to the mechanical mode, thus amplifying the
mechanical zero-point fluctuations and directly enhancing the spin-phonon coupling.
This method allows the spin-mechanical system to be driven from the weak-coupling regime to the strong-coupling regime, and even the ultrastrong coupling regime.  In the dispersive regime,
this method gives rise to a large enhancement of the phonon-mediated spin-spin interactions between distant solid-state spins, typically\emph{ two orders of magnitude larger} than that without modulation. As an example, we show that the proposed scheme can  apply  to generating entangled states of multiple spins with high fidelities even in the presence of large dissipations.
\end{abstract}

\maketitle


Hybrid quantum systems combining completely different physical systems can
realize new functionalities that the individual components can never offer \cite{RevModPhys.85.623,kurizki2015quantum}. The strong coupling
regime of interactions between these different subsystems, where coherent interactions dominate
dissipative processes,  is at the heart of  implementing
more complex tasks such as quantum information processing.
However, couplings between different physical systems are often extremely weak, and strong
coupling has been actively pursued since the birth of hybrid quantum systems.

Recently, interfacing solid-state spins with quantum nanomechanical elements has
attracted great interest \cite{PhysRevLett.110.156402,PhysRevLett.111.227602,PhysRevApplied.5.034010,MacQuarrie2017Cooling,
NCCaiJM,PhysRevLett.113.020503,NCOvartchaiyapong,NPBarfuss,PhysRevB.79.041302,P2010A,
PhysRevLett.117.015502,PhysRevApplied.4.044003,PhysRevLett.121.123604,Arcizet2011A,Kolkowitz1603,Sungkun2012Coherent,PhysRevLett.120.213603,PhysRevX.8.041027,PRRes}.  This hybrid spin-mechanical system takes advantages of the
long coherence time of solid-state spins \cite{Balasubramanian2009Ultralong,NaNBhallamudi,NRMCasola,Doherty2013The,Meyer2007,NRMAtature,NCPfender,Neumann1326,NCYaoNY,PhysRevX.4.031022,
NatPhysDolde,PhysRevLett.109.033604,PhysRevLett.110.243602,PhysRevLett.112.116403,Buluta_2011,Zhu2011Coherent,P2010Quantum,
NPhyCJM,RevModPhys.86.153,PhysRevB.87.144516,PhysRevA.88.012329,PhysRevX.6.041060,PhysRevLett.116.143602,PhysRevX.5.031031,PhysRevApplied.10.024011,li2019interfacing} and the enormous Q factors of nanomechanical oscillators \cite{poot2012mechanical}, and has
wide applications ranging from quantum information processing to quantum sensing \cite{degen2017quantum}.  To construct
a spin-mechanical setup, solid-state spin qubits like nitrogen-vacancy (NV) centers in diamond can couple to nanomechanical oscillators either
through mechanical strain \cite{PhysRevLett.110.156402,PhysRevLett.111.227602,PhysRevApplied.5.034010,
MacQuarrie2017Cooling,NCCaiJM,PhysRevLett.113.020503,NCOvartchaiyapong,NPBarfuss} or via external magnetic field gradients \cite{PhysRevB.79.041302,P2010A,PhysRevLett.117.015502,PhysRevApplied.4.044003,PhysRevLett.121.123604,Arcizet2011A}.  However, none of these existing systems
have reached the strong coupling regime thus far and novel approaches are needed to improve the spin-phonon and the spin-spin interactions
such that they can enter the strong coupling regime.

In this work, we introduce an experimentally feasible and simple approach that can exponentially enhance the spin-phonon, and the phonon-mediated
spin-spin couplings in a spin-mechanical system using only linear resources.
Through modulating the spring constant of the cantilever in time,
we can acquire a tunable and two-phonon drive to the mechanical mode \cite{PhysRevLett.67.699}, thus amplifying the mechanical zero-point fluctuations \cite{PhysRevLett.107.213603,NJPSzorkovszky,NCLemonde,Liao_2014,PhysRevLett.119.053601,yin2017nonlinear}. This amplification directly enhances the spin-phonon magnetic or strain coupling but without the need to use additional nonlinear resources \cite{PhysRevLett.114.093602,SRLi2016,PhysRevLett.120.093601,PhysRevLett.120.093602}. Thus, this proposal  could implement nonlinear processes with only linear resources, and significantly simplifies the
experimental realization.
We show that the spin-mechanical system can be driven from the weak-coupling regime to the strong-coupling regime,
and even the ultrastrong-coupling regime.
When considering multiple solid-state spins coupled to the same cantilever
in the dispersive regime \cite{PhysRevA.80.022335,PhysRevA.98.052346},  this method gives rise to a large enhancement of the
spin-spin interactions between different spins, typically \emph{two orders of magnitude stronger} than that without spring constant modulation.
As an intriguing application, we show how this approach allows one to generate spin squeezed states with high qualities even in the presence of large dissipations. The proposed method is general, and can apply to other defect centers or
solid-state systems  coupled to a quantum nanomechanical element. Related approaches using bosonic parametric driving for spin squeezing have been considered in the context of  trapped ions \cite{ge2019trapped} and cavity QED \cite{groszkowski2020heisenberg}. This work
differs fundamentally from these proposals with a markedly different kind of hybrid spin-mechanical system.

\begin{figure}
\includegraphics[width=8cm]{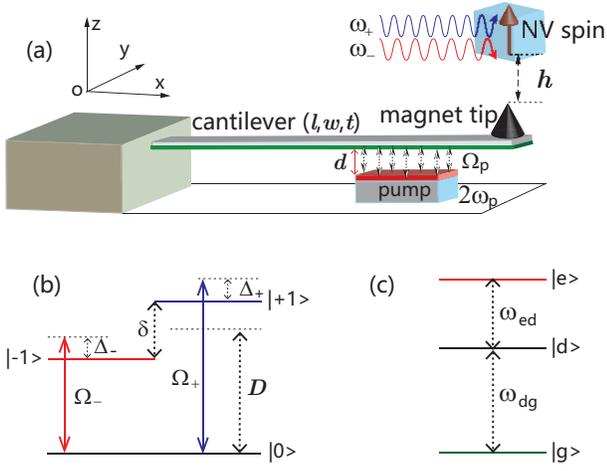}
\caption{\label{fig:wide}(Color online)
(a) Schematic of a single NV center magnetically coupled to a silicon cantilever mechanical resonator, whose spring constant is
periodically  modulated.
The NV center is set just on the top of the corresponding magnet tip with a distance $h\sim25$ nm.
(b) Level diagram of the driven NV center electronic ground states $|m_{s}=0,\pm1\rangle$.
(c) Level diagram of the dressed spin basis states.  }
\end{figure}

\emph{The setup.---}
We consider the spin-mechanical setup, as illustrated in Fig.~1, where
a single NV center is magnetically coupled to the mechanical motion of a cantilever with dimensions $(l,w,t)$, via
a sharp magnet tip attached to its end.
By applying a periodic drive to modulate the spring constant of the cantilever \cite{PhysRevLett.67.699},
the zero-point fluctuations of the mechanical motion can be amplified.
This effect can be realized experimentally by positioning an electrode near the lower surface of the cantilever
and applying a tunable and time-varying voltage to this electrode \cite{PhysRevLett.67.699}.
The gradient of the electrostatic force from the electrode has the effect of modifying the spring constant \footnote{See Supplemental Material for more details}.

For single NV centers, the ground-state energy level structure is shown in Fig.~1(b), with
the ground triplet states $|m_{s}=0,\pm 1\rangle$,
and the zero-field splitting $D=2\pi \times 2.87$ GHz between the degenerate
sublevels $|m_{s}=\pm 1\rangle$ and $|m_{s}=0\rangle $.
We apply a homogeneous static magnetic field $B_{\text{static}}$
to remove the degenerate states $|m_{s}=\pm 1\rangle$ with the Zeeman
splitting $\delta=2g_{e}\mu_{B}B_{\text{static}}$,
where $g_{e}\simeq 2$ and $\mu _{B}=14\ \text{MHz}/\text{mT}$ are the NV's Land\'{e}
factor and Bohr magneton, respectively. We further  apply dichromatic microwave classical fields $B_{x}^{\pm}(t)=B_{0}^{\pm}\cos(\omega_{\pm}t+\phi_{\pm})$ polarized in the $x$ direction
to drive the transitions between the states $|0\rangle$ and $|\pm1\rangle$.
In the rotating frame with the microwave frequencies $\omega_{\pm}$, we obtain the Hamiltonian
$\hat{H}_{\text{NV}}=\sum_{j=\pm1}-\Delta_{j}|j\rangle\langle j|+\frac{\Omega_{j}}{2}(|0\rangle\langle j|+|j\rangle\langle 0|)$,
where $\Delta_{\pm}\equiv|D-\omega_{\pm}\pm\delta/2|$ and $\Omega_{\pm}\equiv g_{e}\mu_{B}B_{0}^{\pm}/\sqrt{2}$. In the following,
we restrict the  discussion to symmetric conditions: $\Delta_{\pm}=\Delta$ and $\Omega_{\pm}=\Omega$.

The Hamiltonian for the nanomechanical resonator with a modulated spring is
$\hat{H}_{\text{mec}}= \frac{\hat{p}_{z}^{2}}{2M}+\frac{1}{2}k(t)\hat{z}^{2}$,
where $\hat{p}_{z}$ and $\hat{z}$ are the cantilever's momentum and displacement operators,
with  effective mass $M$ and fundamental frequency $\omega_{m}$.  The spring
constant of the cantilever is modified (pumped) at a frequency $2\omega_p$ by the electric field from the capacitor plate, $k(t)=k_{0}+k_{r}(t)$, where
$k_{0}=M \omega_{m}^{2}$ is the fundamental spring constant, and the time-dependent correction item
$k_{r}(t)=\partial F_{e}/\partial z=\Delta k \cos(2\omega_{p}t)$ \cite{Note1}. Here, $F_{e}=\partial (C_r V^{2})/(2\partial z)$
is the tunable electrostatic force exerted on the cantilever by the electrode \cite{PhysRevLett.67.699}, with
$C_r$  the electrode-cantilever capacitance, and $V(t)$  the time-dependent voltage, which is assumed to have the form
$V(t)=V_0+V_p\cos2\omega_pt$. Then, we can obtain  $\Delta k=(\partial^2 C_r/\partial z^2)V_0V_p$.
Expressing the momentum operator $\hat{p}_z$ and the displacement operator $\hat{z}$ with the
oscillator operator $\hat{a}$ of the fundamental oscillating mode and
the zero field fluctuation $z_{\text{zpf}}=\sqrt{\hbar/2M\omega_{m}}$, i.e., $\hat{p}_z=-i(M\hbar\omega_m/2)^{1/2}(\hat{a}-\hat{a}^{\dag})$ and $\hat{z}=z_{\text{zpf}}(\hat{a}^{\dag}+\hat{a})$,
we obtain ($\hbar =1$)  \cite{Note1}
\begin{eqnarray}
\hat{H}_{\text{mec}}&=&\omega_{m}\hat{a}^{\dag}\hat{a}- \Omega_{p}\cos(2\omega_{p}t)(\hat{a}^{\dag}+\hat{a})^{2},
\end{eqnarray}
where $\Omega_{p}=-\Delta k z_{\text{zpf}}^{2}/2$ is the  classical drive amplitude.

The Hamiltonian $\hat{H}_{\text{int}}=\mu_{B}g_{e}G_{m}\hat{z}\hat{S}_{z}$
describes the magnetic interaction between the NV spin and the cantilever's vibrating mode,
with $G_{m}$ the magnetic field gradient.
We switch to the dressed state basis $\{|d\rangle=1/\sqrt{2}(\vert+1\rangle-\vert-1\rangle)$, $|g\rangle=\cos\theta|0\rangle-\sin\theta|b\rangle$, $|e\rangle=\cos\theta|b\rangle+\sin\theta|0\rangle\}$, with
$|b\rangle=(|+1\rangle+|-1\rangle)/\sqrt{2}$, and $\tan(2\theta)=-\sqrt{2}\Omega/\Delta$.
We assume that the transition frequency between the
dressed states $\vert g\rangle$ and $\vert d\rangle$ becomes comparable with
the oscillator frequency, i.e., $\omega_{dg}\sim \omega_m$.
The total Hamiltonian for this hybrid system under the rotating-wave approximation by dropping the high frequency oscillation and the constant items can be simplified as  \cite{Note1}
\begin{eqnarray}\label{ME1}
\hat{H}_{\text{Total}}&\simeq& \delta_{m}\hat{a}^{\dag}\hat{a}+\frac{\delta_{dg}}{2}\hat{\sigma}_{z}-\frac{\Omega_{p}}{2}(\hat{a}^{\dag 2}+\hat{a}^{2})\notag\\
&+&\lambda(\hat{a}^{\dag}\hat{\sigma}_{-}+\hat{a}\hat{\sigma}_{+}),
\end{eqnarray}
where the coefficients are $\delta_{m}=\omega_{m}-\omega_{p}$, $\delta_{dg}=\omega_{dg}-\omega_{p}$, $\lambda=-\mu_{B}g_{e}G_{m}z_{\text{zpf}}\sin\theta$,
 $\hat{\sigma}_{z}\equiv(|d\rangle \langle d|-|g\rangle\langle g|)$,
$\hat{\sigma}_{+}\equiv|d\rangle\langle g|$, and $\hat{\sigma}_{-}\equiv|g\rangle\langle d|$ \cite{PhysRevB.79.041302}.
Note that the above model Hamiltonian can also be realized for the case where NV spins are coupled to a nanomechanical cantilever via mechanical
strain \cite{Note1}.

\emph{Enhancing the spin-phonon interaction.---}
Considering the Hamiltonian (\ref{ME1}),
we can diagonalize the mechanical part of $\hat{H}_{\text{Total}}$
by the unitary transformation $\hat{U}_{s}(r)=\exp[r(\hat{a}^{2}-\hat{a}^{\dagger2})/2]$,
where  the squeezing parameter $r$ is defined via the relation $\tanh2r=\Omega_{p}/\delta_{m}$.
Then, we can obtain the Rabi Hamiltonian in this squeezed frame \cite{Note1}
\begin{eqnarray}\label{ME2}
\hat{H}_{\text{Rabi}}^{S}=\Delta_{m}\hat{a}_s^{\dag}\hat{a}_s+\frac{\delta_{dg}}{2}\hat{\sigma}_{z}+\lambda_{\text{eff}}(\hat{a}^{\dag}+\hat{a})(\hat{\sigma}_{+}+\hat{\sigma}_{-}).
\end{eqnarray}
Here,  $\Delta_{m}=\delta_{m}/\cosh 2r$. We have neglected the undesired
correction to the ideal Rabi Hamiltonian in the large amplification regime.  This term (with coefficient $\lambda e^{-r}/2$) is explicitly suppressed when we increase the squeeze parameter $r$,
and  is negligible in the large amplification regime $1/e^{r}\sim 0$.
More importantly, we can obtain the exponentially enhanced spin-phonon coupling strength
$\lambda_{\text{eff}}=\lambda e^{r}/2$,
which can be  orders of magnitude
larger than the original coupling strength as shown in Fig.~2(a), and  comparable with $\Delta_{m}$ and $\delta_{dg}$, or even stronger than both of them.

\begin{figure}[t]
\includegraphics[width=8.5cm]{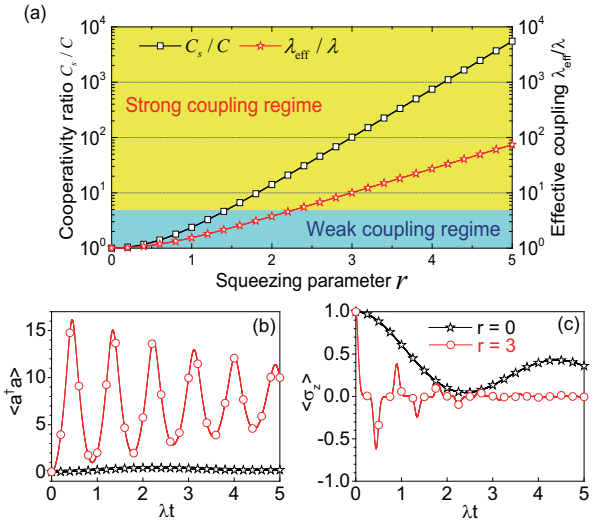}
\caption{\label{fig:wide}(Color online) (a) Cooperativity enhancement $C_{S}/C$ and the spin-phonon coupling enhancement $\lambda_\text{eff}/\lambda$  versus the squeezing parameter $r$, with $\delta_{m}=10\lambda$.  (b) and (c) Quantum dynamics of the mechanical resonator and NV spin with the two-phonon drive ($r=3$ ) or without it ($r=0$ ), where
the parameters are $\delta_{m}=\lambda$, $\delta_{dg}=0$, $\Gamma^S_{m}=\gamma_{NV}=0.1\lambda$.  The mechanical mode and NV center are initially prepared in the vacuum state $\vert 0\rangle_m$ and dressed state $|d\rangle$, respectively.}
\end{figure}
To quantify the enhancement of the spin-phonon coupling \cite{PhysRevLett.75.553}, we exploit the cooperativity $C=\lambda^{2}/\Gamma_{m}\gamma_{\text{NV}}$.
Here, $\Gamma_m$ and $\gamma_{\text{NV}}$ correspond to the effective mechanical dissipation rate
and the dephasing rate of the spin, respectively.
To circumvent the detrimental effect of  amplified mechanical noises, a possible strategy is to use the dissipative squeezing approach \cite{Wollman952,PhysRevLett.115.243601,NCLemonde}, in which  an additional optical  or microwave mode is added to the system, and is used as an engineered reservoir to keep the Bogoliubov mode in its
ground state \cite{Note1}.
This steady-state technique has already been implemented experimentally \cite{Wollman952,PhysRevLett.115.243601,NCLemonde}.
In this case, the squeezed phonon mode equivalently interacts with the thermal vacuum
reservoir, and we can obtain the  master equation  in the squeezed frame  \cite{Note1}
$\dot{\hat{\rho}}=i[\hat{\rho},\hat{H}_{\text{Rabi}}^{S}]+\Gamma_{m}^{S}\mathcal{D}[\hat{a}]\hat{\rho}+\gamma_{\text{NV}}\mathcal{D}[\hat{\sigma}_{z}]\hat{\rho}$,
where $\Gamma_{m}^{S}$ is the engineered effective dissipation rate resulting from the coupling of the mechanical mode to the auxiliary bath.
Therefore, we can also define the effective cooperativity $C_{S}=\lambda_\text{eff}^{2}/\Gamma_{m}^{S}\gamma_{\text{NV}}$.

In Fig.~2(a) we plot the cooperativity enhancement $C_{S}/C\sim e^{2r}/4$, as well as the spin-phonon coupling enhancement $\lambda_{\text{eff}}/\lambda$, versus the squeezing parameter $r$. We find that increasing the parameter $r$ enables an exponential enhancement in the spin-phonon coupling, thus directly giving rise to the cooperativity enhancement.
Figures 2(b,c) show the quantum dynamics of the spin-mechanical system for the cases when
the spring constant is modulated or not. As the spring constant is modulated, the system can
be pumped and driven from the weak-coupling regime to the strong-coupling, or even the ultrastrong-coupling regime.

\emph{Enhancing the phonon-mediated spin-spin interaction.---}
We now consider multiple NV spins coupled to the cantilever through either magnetic or strain coupling.
When the spring constant of the cantilever is modulated, we can obtain the following
Hamiltonian describing the coupled system \cite{Note1}
\begin{eqnarray}
\hat{H}_{\text{Rabi}}^{N}&=&\Delta_{m}\hat{a}^{\dag}\hat{a}+\sum_{j=1}^{N}[\frac{\delta_{dg}^{j}}{2}\hat{\sigma}_{z}^{j}
+\lambda_{\text{eff}}^{j}(\hat{a}^{\dag}+\hat{a})\hat{\sigma}_{x}^{j}]
\end{eqnarray}
In the following,  we set $\delta_{dg}^{j}=0$ for simplicity.
We apply the unitary polaron transformation $\hat{U}=e^{-i\hat{Z}}$ to $\hat{H}_{\text{Rabi}}^{N}$,
with $\hat{Z}=i\sum_{k=1}^{N}\eta_{k}(\hat{a}^{\dag}-\hat{a})\hat{\sigma}_{x}^{k}$ and the Lamb-Dicke condition $\eta_{k}=\lambda_{\text{eff}}^{k}/\Delta_{m}\ll 1$. In this case, the phonons are only virtually excited and can
mediate effective interactions between the otherwise
decoupled solid-state spins \cite{PhysRevLett.110.156402,PhysRevLett.117.015502}.
Then we can obtain the effective spin-spin interactions \cite{Note1}
$\hat{H}_\text{eff}=\sum_{j,k=1}^{N}\Lambda^{jk}\hat{\sigma}_{x}^{j}\hat{\sigma}_{x}^{k},$
where $\Lambda^{jk}=(1+\exp{4r})\frac{\lambda^{j}\lambda^{k}}{8\delta_{m}}$
is the effective coupling strength between the $j$th and the $k$th NV spins via the exchange of virtual phonons.
Here the effective coupling strength for the phonon-mediated spin-spin interactions has an amplification factor $e^{4r}$, and can be orders of magnitude larger than that without mechanical amplification. In the case of homogeneous coupling, we have
\begin{eqnarray}\label{OAT}
\hat{H}_{\text{OAT}}&=&\Lambda\hat{J}_{x}^2,
\end{eqnarray}
where $\Lambda=(1+e^{4r})\frac{\lambda^2}{8\delta_{m}}$, and $\hat{J}_{x}=\sum_{j=1}^{N}\hat{\sigma}_{x}^{j}$.
This Hamiltonian corresponds to the one-axis twisting
interaction \cite{kitagawa1993squeezed} or equivalently belongs to the well-known Lipkin-Meshkov-Glick (LMG) model \cite{lipkin1965validity,zhou2017simulating}.

\begin{figure}
\includegraphics[width=9cm]{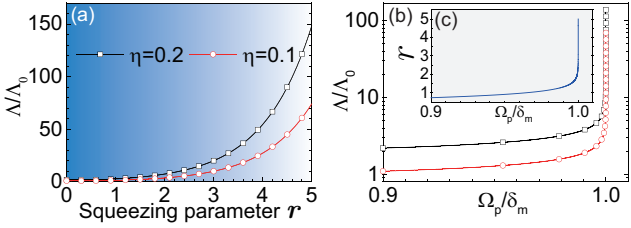}
\caption{\label{fig:wide}(Color online)  The enhanced effective spin-spin coupling strength $\Lambda$ varying with the squeezing parameter $r\in[0,5]$ (a) and the pump amplitude $\Omega_{p}/\delta_{m}\in[0.9,1)$ (b), with different constraint $\eta=0.2$ and $\eta=0.1$. (c) The squeezing parameter $r$ versus $\Omega_{p}/\delta_{m}$. Here we assume that the effective spin-spin coupling strength without the two-phonon drive ($r=0$) is $\Lambda_{0}\simeq0.1\lambda$. }
\end{figure}
Figure~3 shows the ratio of the enhanced spin-spin coupling strength $\Lambda$ to the bare coupling $\Lambda_0$ as a function
of the parameter $r$. Increasing the mechanical parametric  drive gives rise to a large enhancement of the phonon-mediated spin-spin interaction,
typically two orders of magnitude  larger than the bare coupling. Note that since the phonon modes have been adiabatically eliminated, this amplified spin-spin coupling does not rely on the specific frame of phonons.
This large, controllable phonon-mediated interaction
between NV spins is at the heart of realizing many quantum technologies such as quantum computation and simulation.

\emph{Applications.---} We now consider generating entangled states with this setup in the presence of  dissipations.
Here, we focus on entangling multiple separated NV spins through exchanging virtual phonons \cite{Note1}.
The one-axis twisting Hamiltonian (\ref{OAT}) can be
used to produce spin squeezed states which generally exhibit many-body entanglement.
Taking into account the effect of spin dephasing, the system is described by the following master equation
$\dot{\hat{\rho}}=i[\hat{\rho},\hat{H}_{\text{OAT}}]+\sum_{j=1}^N\gamma_{\text{NV}}\mathcal{D}[\hat{\sigma}_{z}^{j}]\hat{\rho}.$
Here, we investigate  the metrological spin squeezing parameter   $\xi ^{2}_R$, the spin squeezing
parameter $\xi ^{2}_S$ \cite{ma2011quantum}, and the metrological gain (the gain of phase sensitivity relative to the standard
quantum limit) $(\Delta\theta_{\text{SQL}}/\Delta\theta)^{2}$ \cite{pezze2018quantum}. When $\xi ^{2}_{S/R}<1$, the states can be shown to be entangled, and have direct implications for spin ensemble-based metrology applications ($(\Delta\theta_{\text{SQL}}/\Delta\theta)^{2}>1$) \cite{pezze2018quantum}.

Figures 4(a,b) show the time evolution of the spin squeezing parameter $\xi ^{2}_{S/R}$ and metrological gain under different $r$.
For a fixed interaction time and in the presence of  spin dephasing,  the spin squeezing parameter $\xi ^{2}_{S/R}$  and metrological gain can be improved significantly by increasing $r$.  Without mechanical amplification, the spin-squeezed state  is seriously spoiled by the detrimental decoherence. However, when modulating the spring constant of the mechanical cantilever and increasing the pump amplitude $\Omega_p$ to a critical value, the quality of the produced state and the speed for generating it can be greatly improved.

\begin{figure}
\includegraphics[width=9cm]{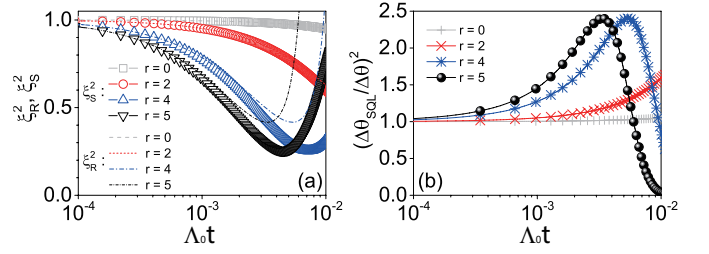}
\caption{\label{fig:wide}(Color online) (a) Metrological spin squeezing parameter $\xi_{R}^{2}$ and spin squeezing parameter $\xi_{S}^{2}$, and (b)  metrological gain $(\Delta\theta_{\text{SQL}}/\Delta\theta)^{2}$ versus time under different values for the mechanical
amplification  parameter $r$, with the initial state of the NV spins $|g\cdots g\rangle$ and the coefficients $\delta_{dg}=0$, $\Lambda_{0}\simeq 0.1\lambda$, $\gamma_{NV}=0.001\lambda$, and $N=6$. }
\end{figure}

\emph{Experimental  feasibility.---}
To examine the feasibility of this proposal for experiments, we
 consider a silicon cantilever with dimensions $(l=6, w=0.1, t=0.05)$ $\mu$m.
The fundamental frequency and the zero-field fluctuation can be expressed as
$\omega_{m}\sim3.516\times(t/l^{2})\sqrt{E/12\varrho}\sim2\pi\times11$ MHz (with its quality factor $Q$ about $10^{5}$--$10^{6}$) and $z_{\text{zpf}}=\sqrt{\hbar/2M\omega_{m}}\sim 2.14\times10^{-13}$ m,
with Young's modulus $E\sim 1.3\times10^{11}$ Pa, the mass density $\varrho\sim 2.33\times10^{3}$ $\text{kg}/\text{m}^{3}$, and effective mass $M \sim\varrho lwt/4$. Assuming an environmental
temperature 10 mK in a dilution refrigerator, the thermal
phonon number is about  $n_{th}\sim100$. Thus the effective  mechanical dissipation rate is $\Gamma_m=n_{th}\omega_{m}/Q\sim 2\pi\times 1$kHz. It is worth noting  that the strain coupling scheme is particularly suitable for the case of multiple NV centers simultaneously coupled to the same cantilever.
For the case of magnetic coupling, we assume that the magnetic tip has a transverse width of $50$ nm, longitudinal height of $100$ nm,
and a radius of curvature of the tip $\sim20$ nm. An array of NV
centers are placed homogeneously and sparsely in the vicinity
of the upper surface of the diamond sample, just under the
magnet tips one by one with the same distance
 $h\sim25$ nm. Note that individual, optically resolvable
NV centers can be implanted determinately  at a single spot 5-10 nm below the surface
of the diamond sample by targeted ion implantation \cite{Kolkowitz1603,Sungkun2012Coherent}, in direct analogy to the excellent control over the locations and distances between the ions in trapped ions.

 In order to ensure that
the magnetic dipole interactions between adjacent centers can
be ignored, we assume that the distance between the adjacent
NV centers (or the adjacent magnetic tips) is about 50 nm.
Furthermore, the distance between the adjacent magnetic tips
and NV centers is also about  50 nm. Therefore, for each
NV spin, the influence caused by the adjacent magnetic tips
can be ignored.
The first-order gradient magnetic field caused by the sharp magnetic tip is about $G\sim1.7\times10^{7}$ $\text{T}/\text{m}$.
We can  obtain the magnetic coupling strength between the cantilever and the NV spin as $\lambda/2\pi\sim100$ kHz.
We expect the variations in the size and spacing of the nanomagnets and NV centers  give rise to a degree of
disorder in the system \cite{PhysRevA.98.052346}.  The disorder  makes  the coupling $\lambda$ cannot be the
same for all of the NV centers. However, as analyzed \cite{PhysRevA.98.052346,Note1}, when the disorder factor
is less than $5\%$, its effective influence on the system can be neglected.

We assume that the  pump  frequency and the amplitude are
respectively $\omega_{p}/2\pi\sim10$ MHz and $\Omega_{p}/2\pi\sim1$ MHz
\cite{RevModPhys.67.249,IMBODEN201489,nnanoLiMo,NCTaoY,RevKLE,Yang2000Surface,APLBrantley}.
In this protocol, the squeezing parameter satisfies $r\in[0,5]$, and then we can obtain the effective spin-phonon coupling $\lambda_{\text{eff}}\sim 100\lambda$ and the effective spin-spin coupling $\Lambda\sim 100\Lambda_{0}\sim 10 \lambda$.
On the other hand, the single NV spin decoherence in diamond
is mainly caused by the coupling of the surrounding electron or nuclear
spins, such as the electron spins P1 centers, the nuclear spins $^{14}\text{N}$ spins and $^{13}\text{C}$ spins.
With the development of the dynamical decoupling
techniques \cite{Bar2013Solid,NatDuJf,PhysRevB.83.081201,PhysRevB.85.115303,PhysRevLett.101.180403,
PhysRevLett.105.200402}, the dephasing time for a single NV center in diamond is about $T_{2}\sim 1/\gamma_{NV}\sim 1$ ms.
Based on the above parameters, we have the magnified cooperativity $C_{S}> 10^6$ with this spin-mechanical hybrid system, much larger than that (about $C_S>10^2$) achieved in a cavity QED or circuit QED system \cite{PhysRevLett.120.093601,PhysRevLett.120.093602}.

Another issue that should be considered is the noise suppression for this system.  In the presence of the mechanical amplification,
the noise coming from the mechanical bath is also amplified. As discussed above, to circumvent such undesired noises, a possible strategy is to use the dissipative squeezing approach. In order to generate the desired squeezed-vacuum reservoir, the  mechanical mode should be prepared in the squeezed state with the squeezing parameter $r\sim1.5$ through the dissipative squeezing method. Note that recent experiments have already demonstrated the generation of squeezed phonon states with the  squeezing parameter $r\sim1.45$ by dissipative squeezing \cite{Kienzler2015}, which corresponds to a 12 dB reduction below the standard quantum limit.

\emph{Conclusion.---}
In this work, we propose an experimentally feasible and simple scheme for exponentially
enhancing the spin-phonon and the spin-spin interactions in a spin-mechanical system with only linear resources.
We show that, by modulating the spring constant of the mechanical cantilever with a
time-dependent pump, the mechanical zero-point fluctuations can be amplified, giving rise to a large enhancement of the spin-phonon and the phonon-mediated spin-spin interactions. The proposed method is general, and can apply to
other defect centers or solid-state systems such as silicon-vacancy center (SiV),
germanium-vacancy center (GeV), and tin-vacancy center (SnV) in diamond
\cite{PhysRevX.8.021063,PhysRevLett.112.036405,PhysRevLett.113.113602,PhysRevLett.118.223603,PhysRevLett.119.253601} coupled to a
quantum nanomechanical element.

\begin{acknowledgments}
P.B.L is supported by the National Natural Science Foundation of China under
Grant No. 11774285, and Natural Science Basic Research Program of Shaanxi  (Program No. 2020JC-02).
Y.Z is supported by the the Natural Science Foundation
of Hubei Province No. 2020CFB748, and the Doctoral Scientific Research
Foundation of HUAT under Grant No. BK201906.
F.N. is supported in part by the:
MURI Center for Dynamic Magneto-Optics via the
Air Force Office of Scientific Research (AFOSR) (FA9550-14-1-0040),
Army Research Office (ARO) (Grant No. Grant No. W911NF-18-1-0358),
Asian Office of Aerospace Research and Development (AOARD) (Grant No. FA2386-18-1-4045),
Japan Science and Technology Agency (JST) (via the Q-LEAP program, and the CREST Grant No. JPMJCR1676),
Japan Society for the Promotion of Science (JSPS) (JSPS-RFBR Grant No. 17-52-50023, and
JSPS-FWO Grant No. VS.059.18N), the RIKEN-AIST Challenge Research Fund,
the Foundational Questions Institute (FQXi), and the NTT PHI Laboratory. Part of the simulations are coded
in PYTHON using the QUTIP library \cite{Johansson2012QuTiP, Johansson2013QuTiP}.
\end{acknowledgments}


%

\onecolumngrid

\appendix

\clearpage

\section*{Supplemental Material:  }


\setcounter{equation}{0}
\setcounter{figure}{0}
\setcounter{table}{0}
\setcounter{page}{1}
\makeatletter
\renewcommand{\theequation}{S\arabic{equation}}
\renewcommand{\thefigure}{S\arabic{figure}}
\renewcommand{\bibnumfmt}[1]{[S#1]}
\renewcommand{\citenumfont}[1]{S#1}
\begin{quote}
In this Supplemental Material,
we first present more details on realizing the mechanical parametric amplification (MPA)
through modulating the spring constant of the cantilever with a time-dependent pump in this setup.
Second, we derive the total Hamiltonian of this hybrid system and discuss the basic idea of
enhancing the spin-phonon and spin-spin coupling at the single quantum level. Meanwhile, we
show  detailed descriptions and discussions on the validity of the effective Rabi model in this work.
We also discuss one potential strategy for engineering the effective dissipation rate of the mechanical mode.
Third, we discuss two specific applications of the spin-mechanical setup with the proposed method,
i.e., adiabatically preparing Schr\"odinger cat states and entangling multiple separated NV spins via exchanging virtual phonons.
Finally, we present some discussions on applying the basic idea to enhance the strain coupling between the NV spins and the diamond nanoresonator.
\end{quote}

\subsection{Realizing MPA through modulating the spring constant}
In this scheme, in order to realize MPA,
we apply the periodic drive to modulate the spring constant of the cantilever.
This can be accomplished by positioning an electrode near the lower surface of the cantilever and
applying a tunable time-varying voltage.
As shown in Fig.~1(a) in the main text, the electrode materials are homogeneously coated on the lower surface of the cantilever,
and another electrode plate with the tunable oscillating pump is placed just under the cantilever.
The Hamiltonian of this mechanical system with the time-dependent spring constant is
\begin{eqnarray}\label{SME1}
\hat{H}_{\text{mec}}&=&\frac{\hat{p}_{z}^{2}}{2M}+\frac{1}{2}k(t)\hat{z}^{2}=\frac{\hat{p}_{z}^{2}}{2M}+\frac{1}{2}k_{0}\hat{z}^{2}+\frac{1}{2}k_{r}(t)\hat{z}^{2}.
\end{eqnarray}
The gradient of the electrostatic force from
the electrode has the effect of modifying the spring constant according to $k(t)=k_{0}+k_{r}(t)$,
with $k_{0}=\omega_{m}^{2}M$ the unperturbed fundamental spring constant and the time-varying pump item
\begin{eqnarray}\label{SME2}
k_{r}(t)\equiv \partial^{2} (C_r V^{2})/(2\partial \hat{z}^{2})=\partial F_{e}/\partial \hat{z}=\Delta k \cos(2\omega_{p}t).
\end{eqnarray}
Here, $F_{e}=\partial (C_r V^{2})/(2\partial \hat{z})$ is the tunable electrostatic force exerted on the cantilever by
the electrode, $\hat{z}$ is the cantilever displacement, $\Delta k$ is the variation of the spring constant,
and $2\omega_{p}$ is the pump frequency.
Therefore, these two electrode plates form a general parallel-plate capacitor, and its capacitance is $C_r=\varepsilon S/(d+\hat{z})$. Here, $\varepsilon\equiv\varepsilon_{0}\varepsilon_{r}$ is the permittivity, $\varepsilon_{0}$ and $\varepsilon_{r}$ are the vacuum and the relative permittivity, respectively, $S$ is the effective area, and $(d+\hat{z})$ is the distance between the two plates. Here we assume the voltage $V=V_{0}+V_{p}\cos2\omega_{p}t$ with $V_{0}>V_{p}$.
Substituting this into (\ref{SME2}) and keeping only the $2\omega_{p}$ item,
we can obtain the time-varying spring constant
\begin{eqnarray}\label{SME3}
k_{r}(t)\simeq\frac{2V_{0}V_{p}\varepsilon S}{d^{2}}\times\cos2\omega_{p}t.
\end{eqnarray}

\begin{figure}
\label{SFig1}
\includegraphics[width=8cm]{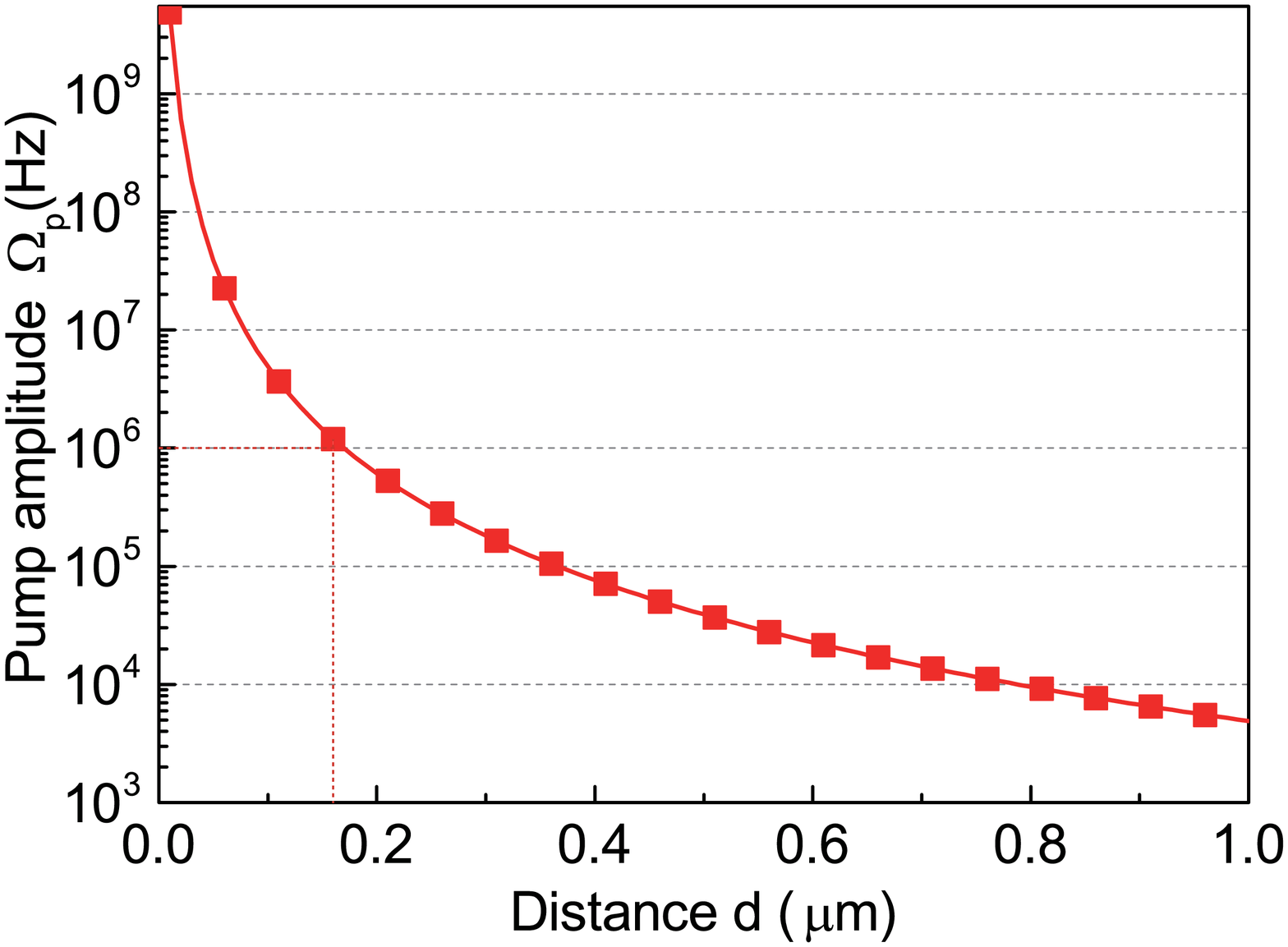}
\caption{\label{fig:wide}(Color online) The pump amplitude $\Omega_{p}$ varying with the distance $d$ between the cantilever's lower surface and the electrode plate. The parameters are set as: the static voltage $V_{0}=10\ V$,
the oscillating voltage $V_{p}=2\ V$, the zero field fluctuation $z_{\text{zpf}}\sim2.14\times10^{-13}m$, and the parallel-plate effective area $S\sim1.0\ \mu\text{m}\times0.1\ \mu$m.  }
\end{figure}
Defining the displacement operator $\hat{z}=z_{\text{zpf}}(\hat{a}^{\dag}+\hat{a})$
with the zero field fluctuation $z_{\text{zpf}}=\sqrt{\hbar/2M\omega_{m}}$,
we can quantize the Hamiltonian of the cantilever $\hat{H}_{\text{mec}}$ ($\hbar =1$),
\begin{eqnarray}\label{SME4}
\hat{H}_{\text{mec}}&=&\omega_{m}\hat{a}^{\dag}\hat{a}- \Omega_{p}\cos(2\omega_{p}t)(\hat{a}^{\dag}+\hat{a})^{2},
\end{eqnarray}
where $\omega_{m}=\sqrt{k_{0}/M}$ is the fundamental frequency, and $\Omega_{p}=-\Delta k z_{\text{zpf}}^{2}/2$ is the nonlinear drive amplitude.
As a result, utilizing this method, we obtain the second-order nonlinear drive through modulating the spring constant in time. As illustrated in Fig.~S1,
we plot this nonlinear amplitude $\Omega_{p}$ varying with the distance $d$ between this two electrode plates.

Note that we can tune the spring constant of this mechanical resonator through modifying $V$, $\varepsilon$, $S$, and $d$. Therefore, we can assume that $\Delta k$ is a time-independent constant ($\partial_{t}\Delta k=0$)
for the case of exponentially enhancing the spin-phonon and spin-spin couplings in this spin-mechanical system.
On the other hand, to ensure the adiabaticity of this dynamical process and to accomplish the adiabatic preparation of the Schr\"odinger cat state,
we can also assume that $\Delta k(t)$ is a slowly time-varying parameter  (which means $\partial_{t}\Delta k\approx0$).
For these two different cases, we will make specific discussions in the following sections.

\subsection{The Hamiltonian for this hybrid system }

The motion of the cantilever attached with the magnet tip produces the time-dependent gradient
magnetic field $\vec{B}(t)=\vec{B}\cos\omega_{m}t$ at the corresponding NV spin,
with $\vec{B}=(B_{x},\ B_{y},\  B_{z})$ the gradient magnetic field vectors
and the cantilever's fundamental frequency $\omega_{m}$.
Because $\omega_{m}$ is much smaller than the energy transition frequency ($\omega_{m}\ll D\pm \delta/2$),
we can ignore the far-off resonant interactions between the NV spin and the
gradient magnetic fields along the $x$ and $y$ directions.
In the rotating frame at the frequency $\omega_{m}$, the Hamiltonian for describing the magnetic interaction
between the mechanical mode and the single NV center is
\begin{eqnarray}\label{SME5}
\hat{H}_{\text{int}}&=&\mu_{B}g_{e}G_{m}\hat{z}\hat{S}_{z}=\lambda_{0}(\hat{a}^{\dag}+\hat{a})\hat{S}_{z},
\end{eqnarray}
where $\lambda_{0}=\mu_{B}g_{e}G_{m}z_{\text{zpf}}$ is the magnetic coupling strength.

Then we apply the dichromatic microwave classical fields $B_{x}^{\pm}(t)$
(with frequencies $\omega_{+}$ and $\omega_{-}$) polarized in the $x$ direction
to drive the transitions between the states $|0\rangle$ and $|\pm1\rangle$.
The Hamiltonian for describing the single NV center driven by the dichromatic microwave fields is $\hat{H}_{\text{NV}}=D\hat{S}_{z}^{2}+\frac{1}{2}\delta\hat{S}_{z}+\mu_{B}g_{e}(B_{x}^{+}(t)+B_{x}^{-}(t))\hat{S}_{x}$,
with the classical periodic driving fields $B_{x}^{\pm}(t)=B_{0}^{\pm}\cos(\omega_{\pm}t+\phi_{\pm})$.
For a single NV center,
we can obtain the Hamiltonian in the rotating frame with the microwave frequencies $\omega_{\pm}$,
\begin{eqnarray}\label{SME6}
\hat{H}_{\text{NV}}&=&\sum_{j=\pm}-\Delta_{j}|j\rangle\langle j|+\frac{\Omega_{j}}{2}(|0\rangle\langle j|+|j\rangle\langle 0|),
\end{eqnarray}
where $\Delta_{\pm}\equiv|D-\omega_{\pm}\pm\delta/2|$ and $\Omega_{\pm}\equiv g_{e}\mu_{B}B_{0}^{\pm}/\sqrt{2}$.
For simplicity, we set $\Delta_{\pm}=\Delta$ and $\Omega_{\pm}=\Omega$ in the following discussions.
The Hamiltonian (\ref{SME6}) couples the state $|0\rangle$ to a ``bright"  state
$|b\rangle=(|+1\rangle+|-1\rangle)/\sqrt{2}$, while the ``dark"  state $|d\rangle=(|+1\rangle-|-1\rangle)/\sqrt{2}$ is decoupled.
The resulting eigenbasis of $\hat{H}_{\text{NV}}$ is therefore given by $|d\rangle$ and the two dressed states
$|g\rangle=\cos\theta|0\rangle-\sin\theta|b\rangle$ and $|e\rangle=\cos\theta|b\rangle+\sin\theta|0\rangle$,
where $\tan(2\theta)=-\sqrt{2}\Omega/\Delta$. Under this dressed basis, we acquire the eigenfrequencies
$\omega_{d}=-\Delta$, and $\omega_{e/g}=(-\Delta\pm\sqrt{\Delta^{2}+2\Omega^{2}})/\sqrt{2}$.
The energy level diagram of the dressed spin states is illustrated in Fig.~1(c) in the main text.
The parameters $\Omega$ and $\Delta$ are adjustable,
and we can get the suitable energy level which is comparable with the frequency $\omega_{m}$.

Therefore, we obtain the total Hamiltonian
\begin{eqnarray}\label{SME7}
\hat{H}_{\text{Total}}&=&\hat{H}_{\text{NV}}+\hat{H}_{\text{mec}}+\hat{H}_{\text{int}}\notag\\
&=&\omega_{m}\hat{a}^{\dag}\hat{a}+\omega_{eg}|e\rangle\langle e|+\omega_{dg}|d\rangle\langle d|+\frac{1}{2}(\hat{a}^{\dag}+\hat{a})(\lambda|g\rangle\langle d|+\lambda^{'}|d\rangle\langle e|+h.c.)-\Omega_{p}\cos(2\omega_{p}t)(\hat{a}^{\dag}+\hat{a})^{2},
\end{eqnarray}
where the parameters are expressed as $\omega_{eg}=\omega_{e}-\omega_{g}$,
$\omega_{dg}=\omega_{d}-\omega_{g}$, $\lambda=-\lambda_{0}\sin\theta$ and $\lambda^{'}=\lambda_{0}\cos\theta$.
Utilizing the unitary transformation $\hat{U}_{0}(t)=e^{-i\hat{H}_{0}t}$
with $\hat{H}_{0}=\omega_{p}(\hat{a}^{\dag}\hat{a}+|e\rangle\langle e|+|d\rangle\langle d|)$,
we can simplify the Hamiltonian for this hybrid system by dropping the high frequency oscillation and the constant items,
\begin{eqnarray}\label{SME8}
\hat{H}_{\text{Total}}&\simeq& \delta_{m}\hat{a}^{\dag}\hat{a}+\frac{\delta_{dg}}{2}\hat{\sigma}_{z}-\frac{\Omega_{p}}{2}(\hat{a}^{\dag 2}+\hat{a}^{2})+\lambda(\hat{a}^{\dag}\hat{\sigma}_{-}+\hat{a}\hat{\sigma}_{+}).
\end{eqnarray}
In this new basis $\{|d\rangle$, $|g\rangle\}$, we define $\hat{\sigma}_{z}\equiv(|d\rangle \langle d|-|g\rangle\langle g|)$,
$\hat{\sigma}_{+}\equiv|d\rangle\langle g|$, and $\hat{\sigma}_{-}\equiv|g\rangle\langle d|$,
with $\delta_{m}=\omega_{m}-\omega_{p}$ and $\delta_{dg}=\omega_{dg}-\omega_{p}$.

\subsection{Enhanced spin-phonon coupling at the single quantum level}

Considering the Hamiltonian (\ref{SME8}),
we can diagonalize the mechanical mode of $\hat{H}_{\text{Total}}$
by the unitary transformation $\hat{U}_{s}(r)=\exp[r(\hat{a}^{2}-\hat{a}^{\dagger2})/2]$,
where the squeezing parameter $r$ is defined via the relation $\tanh2r=\Omega_{p}/\delta_{m}$.
We can obtain the Hamiltonian in the squeezed frame with the form
\begin{eqnarray}\label{SME9}
\hat{H}_{\text{Total}}^{S}=\hat{H}_{\text{Rabi}}^{S}+\hat{H}_{D}^{S},
\end{eqnarray}
where
\begin{eqnarray}\label{SME10}
\hat{H}_{\text{Rabi}}^{S}=\Delta_{m}\hat{a}^{\dag}\hat{a}+\frac{\delta_{dg}}{2}\hat{\sigma}_{z}+\lambda_{\text{eff}}(\hat{a}^{\dag}+\hat{a})\hat{\sigma}_{x},
\end{eqnarray}
\begin{eqnarray}\label{SME11}
\hat{H}_{D}^{S}=\frac{\lambda e^{-r}}{2}(\hat{a}-\hat{a}^{\dag})(\hat{\sigma}_{+}-\hat{\sigma}_{-}).
\end{eqnarray}
In this squeezed frame, $\hat{H}_{\text{Rabi}}^{S}$ is the Hamiltonian for describing the Rabi model, with
$\Delta_{m}=\delta_{m}/\cosh 2r$. In $\hat{H}_{\text{Rabi}}^{S}$,
we can obtain the exponentially enhanced coupling strength $\lambda_{\text{eff}}\approx\lambda e^{r}/2$,
which will be comparable with $\Delta_{m}$ and $\delta_{dg}$, or even stronger than both of them when increasing $r$.
The remaining Hamiltonian $\hat{H}_{D}^{S}$ describes the undesired correction to the ideal Rabi Hamiltonian.
This item (with coefficient $\lambda e^{-r}/2$) is explicitly suppressed when we increase the squeeze parameter $r$,
and it is negligible in the large amplification regime $1/e^{r}\sim 0$.
Therefore, for enhancing the spin-phonon magnetic coupling through MPA,
we can neglect the influence caused by $\hat{H}_{D}^{S}$ in this scheme.

To verify the discussions above, we make numerical simulations and present  the results  in Fig.~S2.
The initial state is chosen as
$|\Psi^{S}(0)\rangle=|0\rangle_{\text{ph}}|g\rangle$ for different types of Hamiltonian $\hat{H}_{\text{Total}}^{S}$ and $\hat{H}_{\text{Rabi}}^{S}$. Here $|0\rangle_{\text{ph}}$ denotes the vacuum state of the phonon modes.
The time evolution of the average phonon numbers $\hat{a}^{\dag}\hat{a}$ and spin population $\hat{\sigma}_{z}$
is displayed in Fig.~S2 (a) and (b).
We find that, in spite of the negative influence caused by $\hat{H}_{D}^{S}$ in $\hat{H}_{\text{Total}}^{S}$,
the dynamical process given by $\hat{H}_{\text{Total}}^{S}$
maintain a high degree of consistency with the standard Rabi model $\hat{H}_{\text{Rabi}}^{S}$.
Therefore, in this work, we have acquired the effective Rabi
type spin-mechanical interaction with the exponentially enhanced coupling strength $\lambda_{\text{eff}}\approx\lambda e^{r}/2$.

\begin{figure}
\includegraphics[width=12cm]{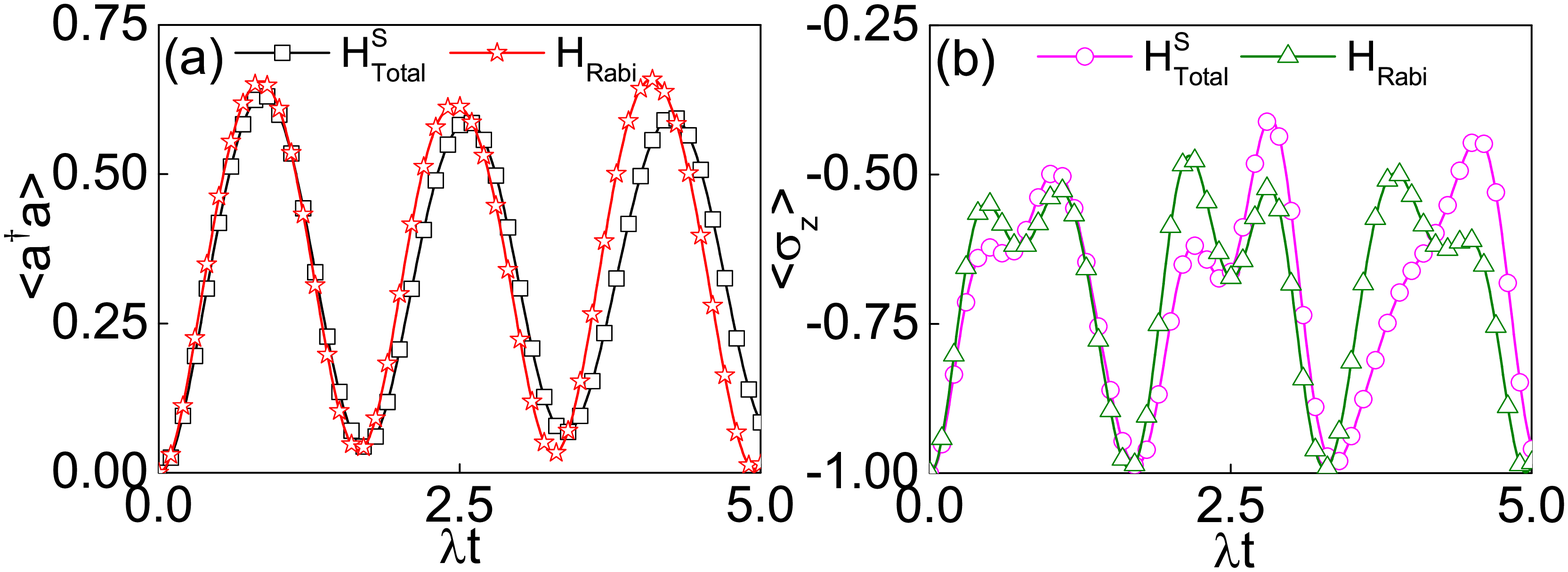}
\caption{\label{fig:wide}(Color online) Dynamical evolution with different Hamiltonian $\hat{H}_{\text{Total}}^{S}$ and $\hat{H}_{\text{Rabi}}^{S}$. (a) Average phonon number $\langle\hat{a}^{\dag}\hat{a}\rangle$, and (b) average spin population $\langle\hat{\sigma}_{z}\rangle$.
Here, the initial state is $|\Psi^{S}(0)\rangle=|0\rangle_{\text{ph}}|g\rangle$,
the coefficients are $\Delta_{m}=20\lambda$ and $\delta_{dg}=2\lambda$, and the squeezing parameter is $r=1.25$.}
\end{figure}

Here we note that, in the presence of parametric amplification,
the noise coming from the mechanical bath is also amplified inevitably.
This adverse factor could corrupt any nonclassical behaviour induced by the enhanced
spin-motion interaction. To circumvent this detrimental effect, a possible strategy is to use the dissipative squeezing method to keep the mechanical mode in its ground state.
Therefore, taking the effective dissipation rate $\Gamma_{m}^{S}$ and the dephasing
rate $\gamma_{\text{NV}}$ into  consideration, in this squeezed frame we can obtain the master equation as follow
\begin{eqnarray}\label{SME12}
\dot{\hat{\rho}}=i[\hat{\rho},\hat{H}_{\text{Rabi}}^{S}]+\Gamma_{m}^{S}\mathcal{D}[\hat{a}]\hat{\rho}+\gamma_{\text{NV}}\mathcal{D}[\hat{\sigma}_{z}]\hat{\rho},
\end{eqnarray}
where $\mathcal{D}[\hat{x}]\hat{\rho}=\hat{x}\hat{\rho}\hat{x}^{\dagger}-\hat{x}^{\dagger}\hat{x}\hat{\rho}/2-\hat{\rho}\hat{x}^{\dagger}\hat{x}/2$. Here we assume that the effective dissipation rate $\Gamma_{m}^{S}$ is
comparable with the dephasing rate $\gamma_{\text{NV}}$ in the following numerical simulations.

By setting the parameters as $\delta_{m}=2\lambda$ and $\delta_{dg}=0$ in Hamiltonian $\hat{H}_{\text{Rabi}}^{S}$,
we plot the time-varying fidelity for the quantum states of
one NV spin ($|g\rangle$ and $|d\rangle$) and the phonon mode ($|n\rangle_\text{ph}$, $n=0,1,2,\cdots $) in Fig.~S3.
Here, the fidelity for the quantum states of NV spin and phonon mode are respectively expressed as
$F_{\text{NV}}^{l}(t)=\langle l|\hat{\rho}_{\text{prace}}^{\text{NV}}(t)|l\rangle^{1/2}$ ($l=g,d$) and
$F_{\text{phonon}}(t)= _\text{ph}\langle n|\hat{\rho}_{\text{prace}}^{\text{phonon}}(t)|n\rangle_\text{ph}^{1/2}$.
We show that, without MAP ( $r=0$) in Fig.~S3(a),
we can obtain the relative weak oscillation curves
for both the NV spin and the mechanical mode.
However, when we increase this parameter from $r=0.5$ to $r=2.0$, corresponding to Fig.~S3(b)-(f),
the amplitude of the time-varying fidelity for the NV spin and phonon mode becomes much larger.
Furthermore, the interval period for these oscillations can also
be substantially shortened with the rate $\sim e^{r}$ when we increase $r$.
These results indicate that,
we can realize the exponentially enhanced strong spin-phonon coupling at the single quantum level in this scheme.

\begin{figure}
\includegraphics[width=12cm]{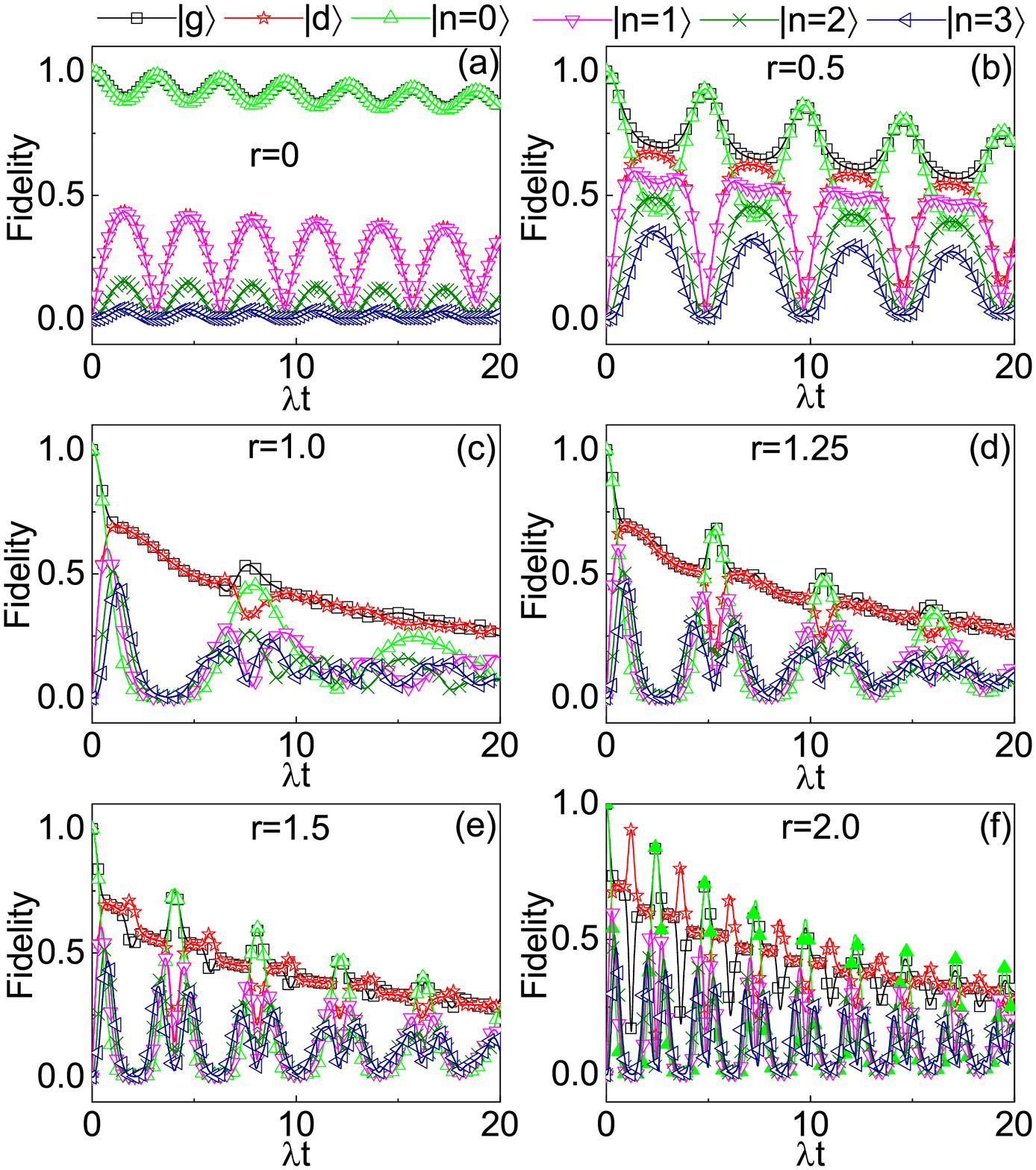}
\caption{\label{fig:wide}(Color online) The dynamical evolution of the fidelity for different states for single NV spin ($|g\rangle$ and $|d\rangle$ correspond to the ground state and excite state)
and the mechanical mode ($|n\rangle_\text{ph}$ stands for the phonon-number state ($n=0,1,2,\cdots$)),
with the coefficients $\delta_{m}=2\lambda$, $\delta_{dg}=0$, and $\gamma_{\text{NV}}=\Gamma_{m}^{S}=0.01\lambda$.
The different squeezed parameters correspond to (a) $r=0$ (no squeezing), (b) $r=0.5$, (c) $r=1$, (d) $r=1.25$,
(e) $r=1.5$, and (f) $r=2.0$.}
\end{figure}

\subsection{Engineering the effective dissipation rate in the squeezed frame}
We note that in the presence of the mechanical amplification,
the noise coming from the mechanical bath is also amplified.
To circumvent this detrimental effect,
a possible strategy is to use the dissipative squeezing approach to
keep the mechanical mode in its ground state in the squeezed frame.
One possible strategy  is to apply an additional optical or microwave mode to this spin-mechanical system,
and utilize it as an ``engineered squeezed reservoir" to keep the mechanical mode in its ground state via dissipative squeezing \cite{PhysRevA.88.063833,PhysRevB.70.205304,PhysRevA.89.013820,PhysRevA.87.033829,PhysRevLett.70.556}.
And this steady-state technique has recently been implemented experimentally \cite{Wollman952,PhysRevLett.115.243601,PhysRevX.5.041037}.
According to the basic idea from the optomechanical system,
we assume this cantilever couples with an additional optical or microwave mode,
and we can describe the coupled system by the Hamiltonian
\begin{eqnarray}\label{OM}
\hat{H}_{\text{OM}}=\Delta_{m}\hat{a}^{\dag}\hat{a}+\omega_{\text{cav}}\hat{c}^{\dag}\hat{c}
-g_{0}\hat{c}^{\dag}\hat{c}(\hat{a}^{\dag}+\hat{a})+(\alpha_{+}e^{-i\nu_{+}t}+\alpha_{-}e^{-i\nu_{-}t})\hat{c}^{\dag}+\text{H.c.}
\end{eqnarray}
In which, $\hat{c}$ ($\hat{a}$) is the photon (phonon) mode annihilation operator, $g_{0}$ is
the optomechanical coupling, $\nu_{\pm}$ and $\alpha_{\pm}$ are the frequency
and amplitude of the two drive lasers, respectively. In the interaction picture, we apply the displacement transformation
$\hat{c}=\overline{c}_{+}e^{-i\nu_{+}t}+\overline{c}_{-}e^{-i\nu_{-}t}+\hat{C}$ into Eq.~(\ref{OM}), with $\overline{c}_{\pm}$ the coherent light field amplitude due to the two
lasers.
Then we can linearize this optomechanical Hamiltonian as
\begin{eqnarray}\label{SME44}
\hat{H}_{\text{IP}}^{\text{L}}=-\hat{C}^{\dag}(D_{+}\hat{a}^{\dag}+D_{-}\hat{a})
-\hat{C}^{\dag}(D_{+}\hat{a}e^{-i2\Delta_{m}t}+D_{-}\hat{a}^{\dag}e^{i2\Delta_{m}t})+\text{H.c.}
\end{eqnarray}
Here, the effective coupling $D_{\pm}=g_{0}\overline{c}_{\pm}$ are strengthened  by the factors $\overline{c}_{\pm}$.
Then we can assume that $D_{+} < D_{-}$ and $D_{\pm}>0$ without loss of generality,
and apply another unitary squeezing operation $\hat{S}^{'}=\exp[\frac{r^{'}}{2}(\hat{a}^{2}+\hat{a}^{\dag2})]$ to Eq.~(\ref{SME44}),
we can get the well known optomechanical cooling Hamiltonian
\begin{eqnarray}\label{SME45}
\hat{H}_{\text{IP}}^{\text{S}}=-\mathcal{O}\hat{C}^{\dag}\hat{a}+\text{H.c.}
\end{eqnarray}
In equation~(\ref{SME45}) above, we have discarded the high frequency oscillation items,
and the relevant definitions are
$\tanh r^{'}=D_{+}/D_{-}$, $\sinh r^{'}=D_{+}/\mathcal{O}$, $\cosh r^{'}=D_{-}/\mathcal{O}$,
and $\mathcal{O}=\sqrt{D_{-}^{2}-D_{+}^{2}}$.
Thus, despite being driven with the classical fields,
this cavity mode acts as a squeezed reservoir leading to mechanical squeezing.
In this new squeezed frame, we can cool the mechanical mode into its ground state $|0\rangle_\text{ph}$,
and in its original frame, this vacuum state corresponds to the squeezed vacuum state $\hat{S}^{'}|0\rangle_\text{ph}$.

On the other hand, in this ancillary photon-phonon interaction system,
the cavity mode at here plays the role of the auxiliary engineered squeezed reservoir,
which can implement an assistance on suppressing the realistic mechanical noise of this cantilever.
For the realistic condition, this cavity is assumed to obey the bad-cavity limit with the large cavity damping rate $\kappa_{C}$,
and its photon state will always stay in the vacuum state.
So we can eliminate this cavity degree $\hat{C}$ and derive the Lindblad master
equation for the reduced density matrix $\hat{\mathbb{\varrho}}$ of the mechanical resonator.
\begin{eqnarray}\label{SME46}
\dot{\hat{\varrho}}=\Gamma_{m}^S(\hat{a}\hat{\varrho}\hat{a}^\dag-\hat{a}^\dag\hat{a}\hat{\varrho}/2-\hat{\varrho}\hat{a}^\dag\hat{a}/2).
\end{eqnarray}
Here, $\Gamma_{m}^S=4\mathcal{O}^{2}/\kappa_{C}$.

Thus, we have accomplished the target of the engineered cavity reservoir, and
 we can suppress the mechanical noise by utilizing the general
squeezed-vacuum-reservoir technique \cite{PhysRevLett.114.093602}.
As a result, this additional cavity mode in this scheme acts as an engineered reservoir
which can cool the mechanical resonator into a squeezed state,
and we can reach the target of engineering the effective mechanical dissipation $\Gamma_{m}^{S}$.

\subsection{Enhancing the phonon-mediated spin-spin interaction}

We consider a row of separated NV centers (the spacing is about $\sim 50$ nm)
magnetically couple to the same mechanical mode of the cantilever, as illustrated  in Fig.~S4.
\begin{figure}
\includegraphics[width=10cm]{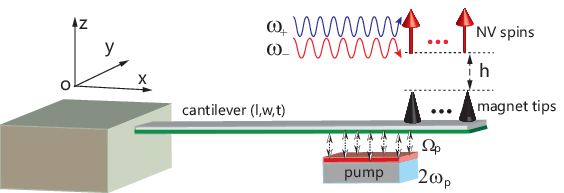}
\caption{\label{fig:wide}(Color online)
 A row of magnet tips are placed at the end of the silicon cantilever.The spring constant of the cantilever is modified (pumped) at frequency
$2\omega_{p}$ by the electric field from the capacitor plate.
Moreover, the amplitude of the pump $\Omega_{p}$ can also be tunable.
Each NV center is set just on top of the corresponding magnet tip with identical distance $h\sim25$ nm.
In addition, two microwave fields polarized in the $x$ direction are applied to drive the NV centers between
the state $|m_{s}=0\rangle$ and the states $|m_{s}=\pm1\rangle$.  }
\end{figure}

According to Eq.~(3) in the main text,
we can obtain the total Hamiltonian,
\begin{eqnarray}\label{SME13}
\hat{H}_{\text{Total}}&\simeq& \delta_{m}\hat{a}^{\dag}\hat{a}-\frac{\Omega_{p}}{2}(\hat{a}^{\dag 2}+\hat{a}^{2})+\sum_{j=1}^{N}[\frac{\delta_{dg}^{j}}{2}\hat{\sigma}_{z}^{j}
+\frac{\lambda^{j}}{2}(\hat{a}^{\dag}\hat{\sigma}_{-}^{j}+\hat{a}\hat{\sigma}_{+}^{j})].
\end{eqnarray}
Applying the same unitary transformation $\hat{U}_{s}(r)$ to $\hat{H}_{\text{Total}}$,
then we can obtain the valid and effective Rabi Hamiltonian by discarding the weak interaction terms in this squeezed frame.
\begin{eqnarray}\label{SME14}
\hat{H}_{\text{Rabi}}^{N}=\Delta_{m}\hat{a}^{\dag}\hat{a}+\sum_{j=1}^{N}[\frac{\delta_{dg}^{j}}{2}\hat{\sigma}_{z}^{j}
+\lambda_{\text{eff}}^{j}(\hat{a}^{\dag}+\hat{a})\hat{\sigma}_{x}^{j}].
\end{eqnarray}
For simplicity, we set $\delta_{dg}^{j}=0$ for each NV spin,
and apply another unitary transformation $\hat{U}=\exp(-i\hat{Z})$ to $\hat{H}_{\text{Rabi}}^{N}$ in Eq.~(\ref{SME14}),
with $\hat{Z}=i\sum_{k=1}^{N}\eta_{k}(\hat{a}^{\dag}-\hat{a})\hat{\sigma}_{x}^{k}$ and $\eta_{k}=\lambda_{\text{eff}}^{k}/\Delta_{m}$.
Here we note that $\eta_{k}$ can be considered as the Lamb-Dicke parameter used in the ion trap system.
We can obtain the effective spin-spin interactions through exchanging the virtual phonons in this spin-mechanical system,
with the constraint $\eta_{k}\ll 1$, which corresponds to $\delta_{m}\gg \lambda e^{3r}/4$.
Through using the  Schrieffer-Wolff transformation $\hat{H}_\text{eff}=\hat{U}\hat{H}_{\text{Rabi}}^{N}\hat{U}^{\dag}$, the mechanical mode can be eliminated from the dynamics.
Then we have the following expressions
\begin{eqnarray}\label{SME115}
\hat{U}(\Delta_{m}\hat{a}^{\dag}\hat{a})\hat{U}^{\dag}=\Delta_{m}\hat{a}^{\dag}\hat{a}
-\sum_{j=1}^{N}\lambda_{\text{eff}}^{j}(\hat{a}^{\dag}+\hat{a})\hat{\sigma}_{x}^{j}
+\sum_{j,k=1}^{N}\frac{\lambda^{j}_{eff}\lambda^{k}_{eff}}{\Delta_{m}}\hat{\sigma}_{x}^{j}\hat{\sigma}_{x}^{k}+O(\eta^{3}).
\end{eqnarray}
\begin{eqnarray}\label{SME116}
\hat{U}[\sum_{j=1}^{N}\lambda_{\text{eff}}^{j}(\hat{a}^{\dag}+\hat{a})\hat{\sigma}_{x}^{j}]\hat{U}^{\dag}=\sum_{j=1}^{N}\lambda_{\text{eff}}^{j}(\hat{a}^{\dag}+\hat{a})\hat{\sigma}_{x}^{j}
-2\times\sum_{j,k=1}^{N}\frac{\lambda^{j}_{eff}\lambda^{k}_{eff}}{\Delta_{m}}\hat{\sigma}_{x}^{j}\hat{\sigma}_{x}^{k}+O(\eta^{3}).
\end{eqnarray}
Keeping only the leading order terms in $\eta_{k}$, we can get the effective Ising type spin-spin interactions,
\begin{eqnarray}\label{SME15}
\hat{H}_\text{eff}&=&\sum_{j,k=1}^{N}\Lambda^{jk}\hat{\sigma}_{x}^{j}\hat{\sigma}_{x}^{k}=\hat{H}_\text{Ising}^{x},
\end{eqnarray}
where $\Lambda^{jk}=\frac{\lambda_{\text{eff}}^{j}\lambda_{\text{eff}}^{k}}{\Delta_{m}}\approx(1+e^{4r})\frac{\lambda^{j}\lambda^{k}}{8\delta_{m}}$
is the effective coupling strength between the $j$th NV spin and the $k$th NV spin.
In the case of homogeneous coupling, we have
\begin{eqnarray}\label{SME16}
\hat{H}_{\text{OAT}}&=&\Lambda\hat{J}_{x}^2,
\end{eqnarray}
where $\Lambda=(1+e^{4r})\frac{\lambda^2}{8\delta_{m}}$, and $\hat{J}_{x}=\sum_{j=1}^{N}\hat{\sigma}_{x}^{j}$.
This Hamiltonian corresponds to the one-axis twisting
interaction or equivalently the well-known Lipkin-Meshkov-Glick (LMG) model. This one-axis twisting Hamiltonian can be
used to produce spin squeezed states which generally exhibit many-body entanglement.
\begin{figure}[t]
\includegraphics[width=8cm]{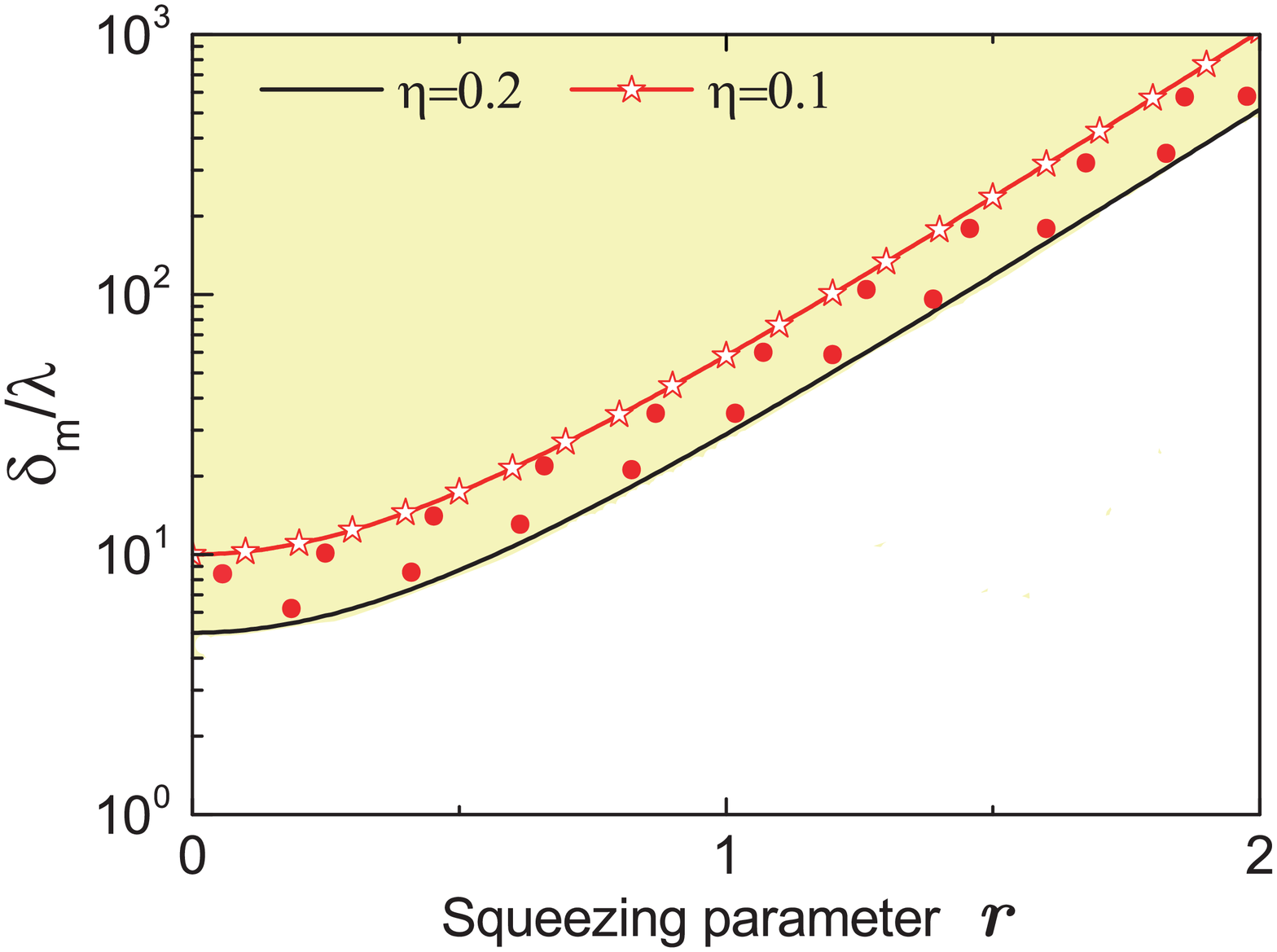}
\caption{\label{fig:wide}(Color online) The constraint value of the parameter $\delta_{m}$ for achieving the effective spin-spin couplings versus the squeezing parameter $r$.
Here, the yellow area shows the valid regime of $\delta_{m}$,
the black solid line and the red solid line with open star represent $\eta=0.2$ and $\eta=0.1$,
the yellow area with red solid dots represents the regime for optimal value of $\delta_{m}$ varying with $r$.}
\end{figure}

When we increase the squeezing parameter $r$, $\lambda_{\text{eff}}$ will be naturally enhanced with a rate $\sim e^{r}$.
Meanwhile, the parameter $\Delta_{m}$ will be reduced with a rate $\sim e^{2r}$.
In order to acquire the indirect spin-spin couplings via the virtual phonon process,
we require $\Delta_{m}\gg \lambda_{\text{eff}}$, which corresponds to the Lamb-Dicke condition $\eta_{k}\equiv\eta\ll 1$.
To obtain the strong spin-spin coupling and ensure the validity of the virtue-phonon process,
we plot the numerical results and find the valid area (the yellow area) in Fig.~S5.
We also point out that the yellow area with red solid dots is the optimal regime for the value of $\delta_{m}$,
which corresponds to the condition $0.1\leq \eta \leq0.2$.

\begin{figure}[t]
\includegraphics[width=12cm]{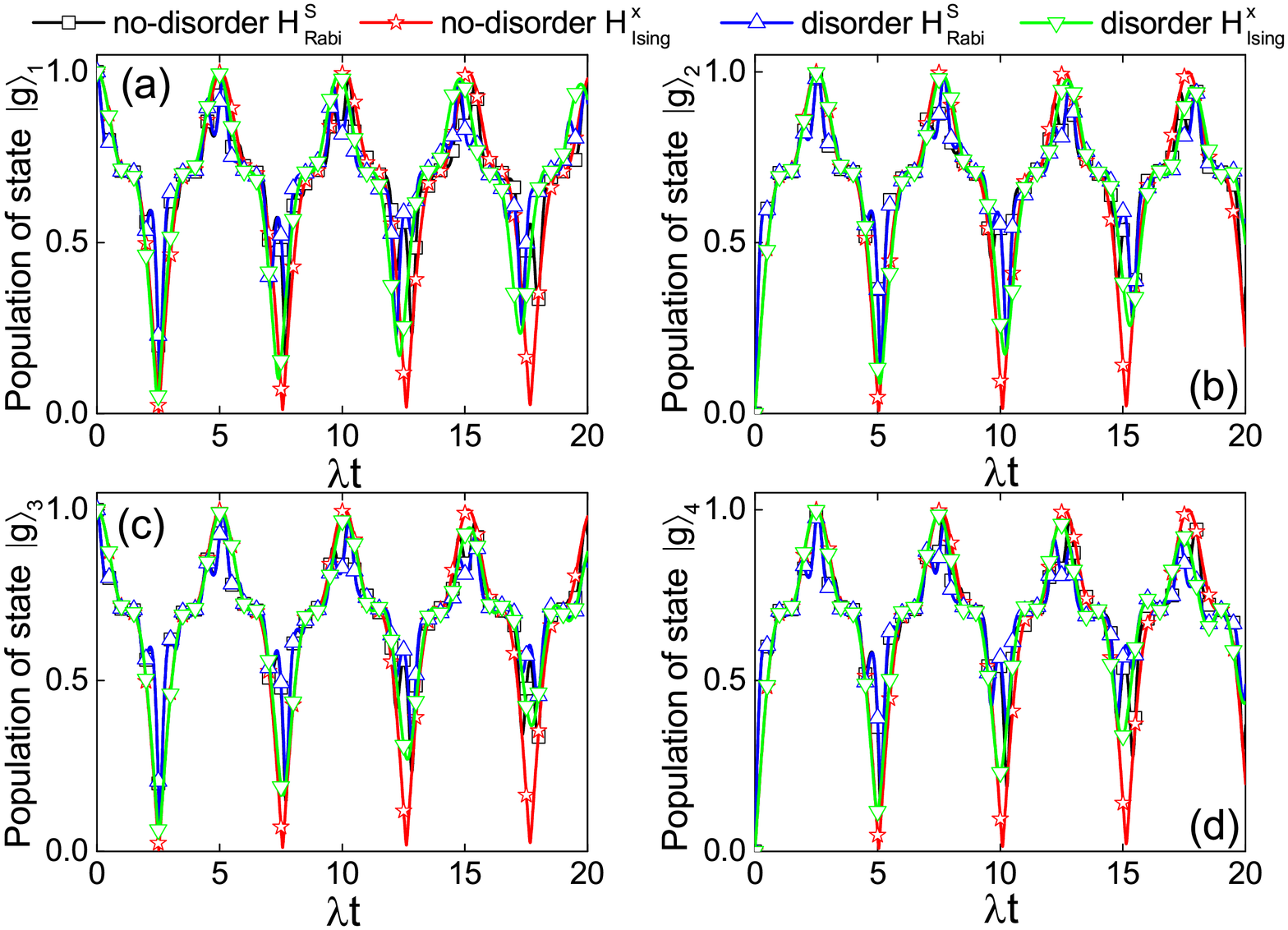}
\caption{\label{fig:wide}(Color online) The comparison of the dynamical population for the ground states $|g\rangle_{j}$ of four different NV centers.
(a) for state $|g\rangle_{1}$, (b) for state $|g\rangle_{2}$, (c) for state $|g\rangle_{3}$, and (d) for state $|g\rangle_{4}$,
with the different Hamiltonian $H_{\text{Rabi}}^{S}$ without disorder (the black solid line with open square),
$H_{\text{Rabi}}^{S}$ with disorder (the blue solid line with open up triangle),
$H_{\text{Ising}}^{x}$ without disorder (the red solid line with open star),
$H_{\text{Ising}}^{x}$ with disorder (the green solid line with open down triangle).
Here, the initial states for this four NV spins and the phonon mode are
$|g\rangle_{1}|e\rangle_{2}|g\rangle_{3}|e\rangle_{4}$ and $|0\rangle_{\text{phonon}}$, respectively,
and the coefficients are $r=1.25$, $\delta_{m}=60\lambda$, and $\gamma_{\text{NV}}^{j}=\Gamma_{m}^{S}=0.01\lambda$.
For homogeneous spins (no disorder), we assume $\delta_{dq}^{j}=0$ and $\lambda^{j}=\lambda$;
while for inhomogeneous spins (disorder), we set the disorder distributions with
$\{\delta_{dq}^{1}=-0.03\lambda,\ \delta_{dq}^{2}=0.03\lambda,\ \delta_{dq}^{3}=0,\ \delta_{dq}^{4}=-0.02\lambda\}$
and $\{\lambda^{1}=1.03\lambda,\ \lambda^{2}=0.98\lambda,\ \lambda^{3}=0.99\lambda,\ \lambda^{4}=1.01\lambda\}$.}
\end{figure}
Taking the effective mechanical dissipation $\Gamma_{m}^{S}$ and the dephasing
rate $\gamma_{\text{NV}}^{j}$ into consideration, we can also write the master equation as follow
\begin{eqnarray}\label{SME17}
\dot{\hat{\rho}}=i[\hat{\rho},\hat{H}_{\text{Rabi}}^{N}]+\Gamma_{m}^{S}\mathcal{D}[\hat{a}]\hat{\rho}+\sum_{j=1}^{N}\gamma_{\text{NV}}^{j}\mathcal{D}[\hat{\sigma}_{z}^{j}]\hat{\rho}.
\end{eqnarray}
In the following numerical simulations, we  assume that the effective mechanical dissipation rate $\Gamma_{m}^{S}$
is comparable with the dephasing rate $\gamma_{\text{NV}}^{j}$.

Here, we take the systematic disorder into consideration for the realistic experimental implementation,
and assume $\delta_{dq}^{j}=0\pm\Delta\delta_{dq}^{j}$ and $\lambda^{j}=\lambda\pm\Delta\lambda^{j}$
are both inhomogeneous.
We constrain the disorder factors $\Delta\delta_{dq}^{j}$ and $\Delta\lambda^{j}$  to less than $\%5$ of $\lambda$.
Under the Hamiltonian $H_{\text{Rabi}}^{N}$ and $H_{\text{Ising}}^{x}$ with different conditions
(homogeneous and inhomogeneous NV spins),
the simulation results for the population in the ground states $|g\rangle_{j}$ for these four NV centers are plotted in Fig.~S6.
We find that, even taking disorder into consideration, the effective Ising Hamiltonian can give rise to the results very close to
those given by the Rabi Hamiltonian.
Based on these results,
we can accelerate the dynamical process exponentially with a rate $\sim e^{4r}$,
which also provides us the most reliable and straightforward evidence for
realizing the exponentially enhanced spin-spin couplings.

\subsection{Preparing the Schr\"{o}dinger cat state adiabatically}
According to the discussions above, we can also assume that
the  amplitude of the pump is a slowly time-varying parameter,
which can be modified slowly enough to ensure the adiabaticity during this dynamical process
(the time-varying rate satisfies $\partial_{t}\Delta k \ll \omega_{m,p}, \ \delta_{m},\ \lambda$).
Then  Eq.~(\ref{SME8}) can be expressed as follow
\begin{eqnarray}\label{SME18}
\hat{H}_{\text{Total}}(t)&\simeq& \delta_{m}\hat{a}^{\dag}\hat{a}+\frac{\delta_{dg}}{2}\hat{\sigma}_{z}-\frac{\Omega_{p}(t)}{2}(\hat{a}^{\dag 2}+\hat{a}^{2})+\frac{\lambda}{2}(\hat{a}^{\dag}\hat{\sigma}_{-}+\hat{a}\hat{\sigma}_{+}),
\end{eqnarray}
where $\Omega_{p}(t)=-\Delta k(t) z_{\text{zpf}}^{2}/2$ is the time-dependent nonlinear amplitude.
Here we can also diagonalize the mechanical mode in the time-dependent Hamiltonian $\hat{H}_{\text{Total}}(t)$
by the similar unitary transformation $\hat{U}_{s}[r(t)]=\exp[r(t)(\hat{a}^{2}-\hat{a}^{\dagger2})/2]$,
with $\tanh2r(t)=\Omega_{p}(t)/\delta_{m}$.
Then we can obtain the total Hamiltonian in this time-varying squeezed frame,
\begin{eqnarray}\label{SME19}
\hat{H}_{\text{Total}}^{S}(t)=\hat{H}_{\text{Rabi}}^{S}(t)+\hat{H}_{D}^{S}(t)+\hat{H}_{V}^{S}(t),
\end{eqnarray}
where
\begin{eqnarray}\label{SME20}
\hat{H}_{\text{Rabi}}^{S}(t)&=&\Delta_{m}(t)\hat{a}^{\dag}\hat{a}+\frac{\delta_{dg}}{2}\hat{\sigma}_{z}+\lambda_{\text{eff}}(t)(\hat{a}^{\dag}+\hat{a})(\hat{\sigma}_{+}+\hat{\sigma}_{-}),
\end{eqnarray}
\begin{eqnarray}\label{SME21}
\hat{H}_{D}^{S}(t)&=&\frac{\lambda e^{-r(t)}}{2}(\hat{a}-\hat{a}^{\dag})(\hat{\sigma}_{+}-\hat{\sigma}_{-}),
\end{eqnarray}
\begin{eqnarray}\label{SME22}
\hat{H}_{V}^{S}(t)&=&\frac{i\dot{r}(t)}{2}(\hat{a}^{2}-\hat{a}^{\dag 2}).
\end{eqnarray}
Here, $\hat{H}_{\text{Rabi}}^{S}(t)$ is the Hamiltonian for describing the time-dependent Rabi model,
with the parameter $\Delta_{m}(t)=\delta_{m}/\cosh 2r(t)$
and the coupling strength $\lambda_{\text{eff}}(t)\approx\lambda e^{r(t)}/2$.
The remaining terms $\hat{H}_{D}^{S}(t)$ and $\hat{H}_{V}^{S}(t)$
describe the undesired corrections to the ideal Rabi Hamiltonian.
Similar to the discussion above, the item $\hat{H}_{D}^{S}(t)$ with
the coefficient $\lambda e^{-r(t)}/2$ is negligible when we increase $r(t)$.
While the other correction term $\hat{H}_{V}^{S}(t)$ with the
coefficient $i\dot{r}(t)/2$ vanishes explicitly with a time-independent ($\dot{r}(t)= 0$) drive amplitude.
Here, we can tune the driving amplitude $\Omega_{p}(t)$ slowly enough to satisfy the relation $\dot{r}(t)\approx 0$.
Therefore, we can also neglect the influence caused by $\hat{H}_{D}^{S}(t)$ during this dynamical process.

To confirm the discussion and analysis above,
we also carry out the numerical simulations, and plot the evolution results in Fig.~S7.
Based on the Hamiltonians $\hat{H}_{\text{Total}}^{S}(t)$ and $\hat{H}_{\text{Rabi}}^{S}(t)$,
the dynamical populations of $\hat{a}^{\dag}\hat{a}$ and $\hat{\sigma}_{z}$ are plotted in Fig.~S7(a) and (b), respectively.
We find that, in spite of the negative influence caused by $\hat{H}_{D}^{S}(t)$ and $\hat{H}_{V}^{S}(t)$,
the dynamical results induced by $\hat{H}_{\text{Total}}^{S}(t)$
are very close to those obtained from the standard Rabi model $\hat{H}_{\text{Rabi}}^{S}(t)$.
Therefore, maintaining the adiabaticity in the whole system,
$\hat{H}_{\text{Rabi}}^{S}(t)$ is still effective and valid to describe
this spin-phonon interaction in this squeezed frame.
\begin{figure}
\includegraphics[width=15cm]{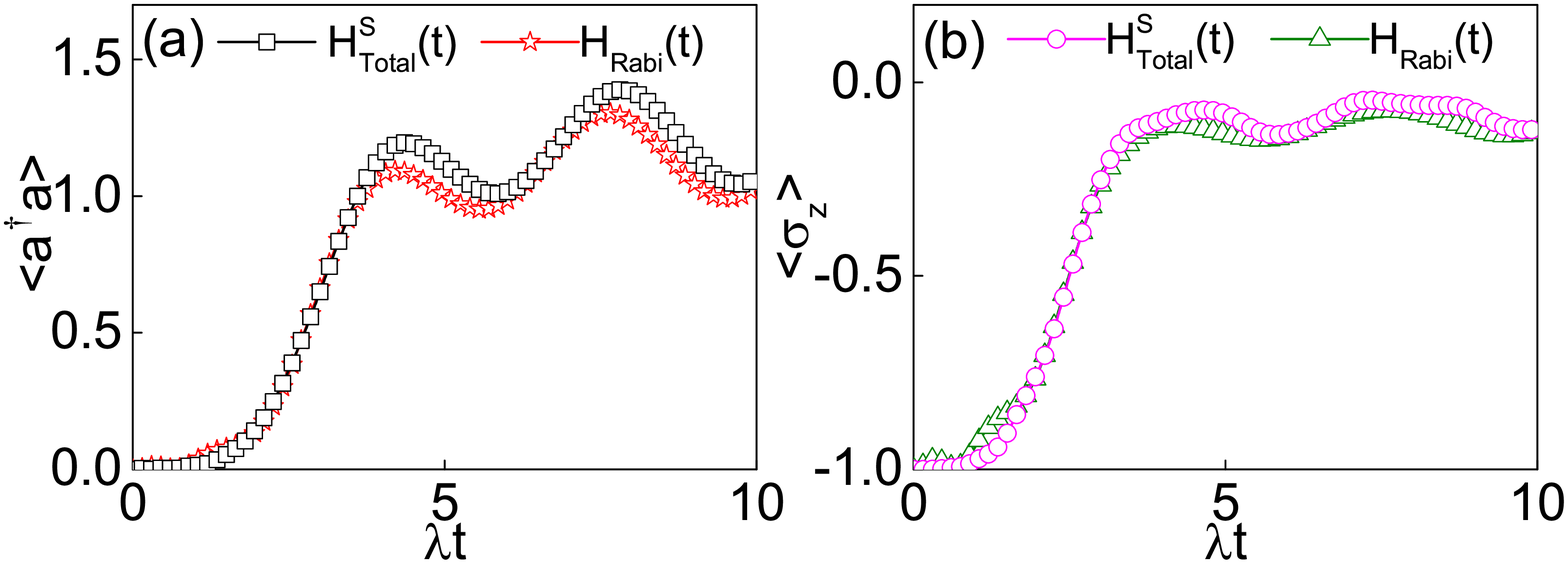}
\caption{\label{fig:wide}(Color online) The dynamical evolution process with different types
of Hamiltonian $\hat{H}_{\text{Total}}^{S}(t)$ and $\hat{H}_{\text{Rabi}}^{S}(t)$. (a) The population of average phonon
number $\langle\hat{a}^{\dag}\hat{a}\rangle$, and (b) the population of the average spin $\langle\hat{\sigma}_{z}\rangle$.
The initial state is $|\Psi^{S}(0)\rangle=|0\rangle_{\text{Phonon}}|g\rangle$,
the coefficients are $\Delta_{m}(t)=\delta_{m}/\cosh[2r(t)]$,
$\delta_{m}=10\lambda$, $\delta_{dg}=0$, $r_{\text{max}}=1.25$,
and the time-dependent squeezing parameter is $r(t)=r_{\text{max}}\tanh(\lambda t/2)$. }
\end{figure}

Considering the Hamiltonian $\hat{H}_{\text{Rabi}}^{S}(t)$ in Eq.~(\ref{SME19}),
we can obtain the time evolution operator as
\begin{eqnarray}\label{SME23}
\mathcal{\hat{U}}_{\text{Rabi}}^{S}(t)=\mathcal{\hat{T}}\exp(-i\int_{0}^{t}\hat{H}_{\text{Rabi}}^{S}(t^{'})dt^{'}),
\end{eqnarray}
where $\mathcal{\hat{T}}$ is the time-ordering operator.
By setting $\delta_{dg}=0$ and utilizing the Magnus expansion,
we can further simplify Eq.~(\ref{SME23}) and have
\begin{eqnarray}\label{SME24}
\mathcal{\hat{U}}_{\text{Rabi}}^{S}(t)&=&e^{[\alpha(t)\hat{a}^{\dag}-\alpha^{*}(t)\hat{a}]\hat{\sigma}_{x}}
e^{-i\Gamma(t,0)\hat{a}^{\dag}\hat{a}},
\end{eqnarray}
where the time-dependent complex parameter is expressed as $\alpha(t)=\frac{\lambda}{2i}\int_{0}^{t}e^{[r(t^{'})-i\Gamma(t,t^{'})]}dt^{'}$,
and another coefficient is $\Gamma(t,t^{'})=\int_{t^{'}}^{t}\Delta_{m}(t^{''})dt^{''}$.
We assume that this spin-mechanical system is initially prepared in the ground state $|\Psi^{S}(0)\rangle=|0\rangle_{\text{Phonon}}|g\rangle$,
and then apply this evolution operator $\mathcal{\hat{U}}_{\text{Rabi}}(t)$ to the initial state $|\Psi^{S}(0)\rangle$.
Finally, we can obtain the target entangled cat state of the single NV spin and the mechanical mode
\begin{eqnarray}\label{SME25}
|\Psi_{\text{Target}}^{S}(t)\rangle=\frac{1}{\sqrt{2}}[|\alpha(t)\rangle|+\rangle_{x}-|-\alpha(t)\rangle|-\rangle_{x}],
\end{eqnarray}
where the states $|\pm\alpha(t)\rangle$ are the  coherent states of the phonon mode,
and the states $|\pm\rangle_{x}$ correspond to the two-level states in the $\hat{\sigma}_{x}$ representation,
with the definition $|\pm\rangle_{x}\equiv(|d\rangle\pm|g\rangle)/\sqrt{2}$.

Therefore,  during this dynamical process with the Rabi interaction $\hat{H}_{\text{Rabi}}^{S}(t)$ from $t=0$ to $t=t_{f}$,
we have acquired an effective adiabatic passage between the ground state
$|0\rangle_{\text{ph}}|g\rangle$ and the target state $|\Psi_{\text{Target}}^{S}(t_{f})\rangle$.
Here we have discarded the negligible adverse factors induced by the $\hat{H}_{D}^{S}(t)$ and $\hat{H}_{V}^{S}(t)$.
To confirm this theoretical analysis and the robustness of this scheme,
we make numerical simulations according to the master equation
\begin{eqnarray}\label{SME26}
\dot{\hat{\rho}}=i[\hat{\rho},\hat{H}_{\text{Total}/\text{Rabi}}^{S}(t)]
+\gamma_{\text{NV}}\mathcal{D}[\hat{\sigma}_{z}]\hat{\rho}+\Gamma_{m}^{S}\mathcal{D}[\hat{a}]\hat{\rho},
\end{eqnarray}
where
$\gamma_{\text{NV}}$ is the dephasing rate,
and $\Gamma_{m}^{S}$ is the effective dissipation rate in the squeezed frame.
We simulate this dynamical evolution with two types of Hamiltonian
$\hat{H}_{\text{Rabi}}^{S}(t)$ and $\hat{H}_{\text{Total}}^{S}(t)$,
and plot the numerical results in Fig.~S8.
We find that, the fidelity reaches  unity when the effective
dispassion and the dephasing rate satisfy $\gamma\leq0.001\lambda$,
while it decreases to $0.97$ when $\gamma\leq0.01\lambda$,
and then decreases to about $0.88$ when $\gamma\leq0.05\lambda$.
Furthermore, the dynamical fidelity obtained from
$\hat{H}_{\text{Rabi}}^{S}(t)$ (the solid line with open symbols)  is in good agreement with that from
$\hat{H}_{\text{Total}}^{S}(t)$ (the solid line with solid symbols).

\begin{figure}
\includegraphics[width=9cm]{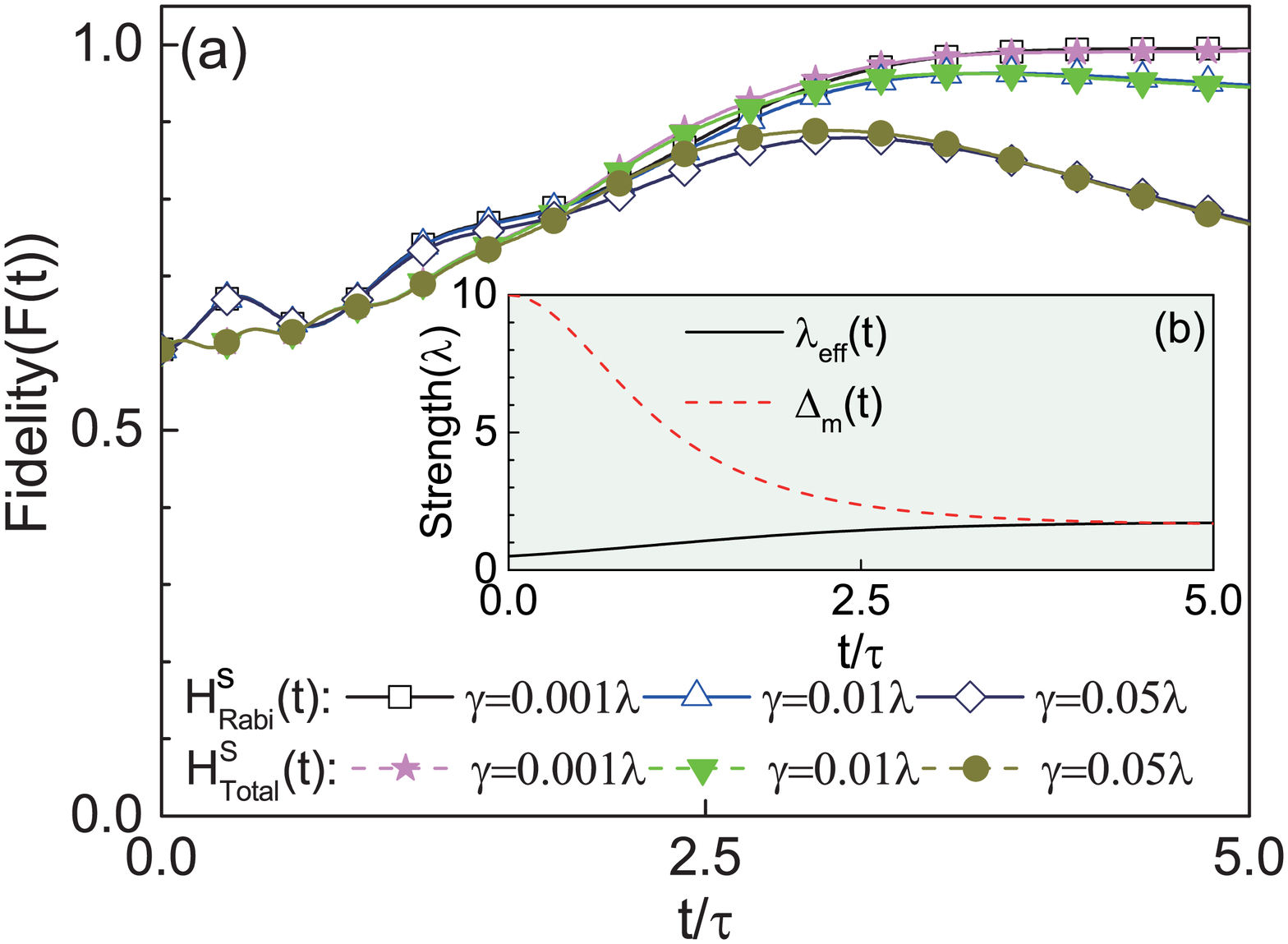}
\caption{\label{fig:wide}(Color online) (a) The fidelity for the target entangled cat state $|\Psi_{\text{Target}}^{S}(t_{f})\rangle$
during the adiabatically dynamical evolution with different types of Hamiltonian
$\hat{H}_{\text{Total}}^{S}(t)$ and $\hat{H}_{\text{Rabi}}^{S}(t)$ and the initial state is $|\Psi^{S}(0)\rangle=|0\rangle_{\text{ph}}|g\rangle$.
(b) The curves for the coefficient $\Delta_{m}(t)$ and the effective coupling $\lambda_{\text{eff}}(t)$ varying with time slowly for adiabaticity.
The coefficients are $\Delta_{m}(t)=\delta_{m}/\cosh[2r(t)]$,
$\delta_{m}=10\lambda$, $\delta_{dg}=0$, $r_{\text{max}}=1.25$,
and the time-dependent squeezing parameter is $r(t)=r_{\text{max}}\tanh[t/(2\tau)]$ with $\tau=1/\lambda$. }
\end{figure}

\subsection{Entangling multiple NV spins}

This spin-mechanical system with exponentially
enhanced coupling strengths could allow us to carry out more complex task:
entangling multiple separated NV spins through exchanging  virtual phonons.
Here we consider $N$ separated NV centers (the spacing is about $50$ nm $\sim0.1\ \mu$m)
magnetically couple to the same mechanical mode of the cantilever.
In this case, according to Eq.~(\ref{SME8}), we can obtain the total Hamiltonian
\begin{eqnarray}\label{SME27}
\hat{H}_{\text{Total}}&\simeq& \delta_{m}\hat{a}^{\dag}\hat{a}-\frac{\Omega_{p}}{2}(\hat{a}^{\dag 2}+\hat{a}^{2})+\sum_{j=1}^N[\frac{\delta_{dg}^{j}}{2}\hat{\sigma}_{z}^{j}
+\frac{\lambda^{j}}{2}(\hat{a}^{\dag}\hat{\sigma}_{-}^{j}+\hat{a}\hat{\sigma}_{+}^{j})].
\end{eqnarray}
Applying the same unitary transformation $\hat{U}_{s}(r)$ to $\hat{H}_{\text{Total}}$,
then we can obtain the effective Rabi Hamiltonian by discarding the weak interaction terms in the squeezed frame
\begin{eqnarray}\label{SME28}
\hat{H}_{\text{eff}}^{S}=\Delta_{m}\hat{a}^{\dag}\hat{a}+\sum_{j=1}^N[\frac{\delta_{dg}^{j}}{2}\hat{\sigma}_{z}^{j}
+\lambda_{\text{eff}}^{j}(\hat{a}^{\dag}+\hat{a})\hat{\sigma}_{x}^{j}],
\end{eqnarray}
where the coefficients are $\Delta_{m}=\delta_{m}/\cosh 2r$, $\delta_{dg}^{j}=\omega_{dg}^{j}-\omega_{p}$ and $\lambda_{\text{eff}}^{j}\approx\lambda^{j}e^{r}/2$.
By setting $\delta_{dg}^{j}=0$ and $\lambda_{\text{eff}}^{j}=\lambda_{\text{eff}}$ for simplicity,
we can obtain the Hamiltonian in the interaction picture with the form
\begin{eqnarray}\label{SME29}
\hat{H}_{\text{IP}}^{S}=\lambda_{\text{eff}}(\hat{a}^{\dag}e^{i\Delta_{m}t}+\hat{a}e^{-i\Delta_{m}t})\hat{J}_{x},
\end{eqnarray}
in which $\hat{J}_{\alpha}\equiv\sum_{j=1}^{N}\hat{\sigma}_{\alpha}^{j}$
are the collective spin operators with $\alpha=\{x,y,z\}$.

The dynamics of the system is governed by the unitary evolution operator $\hat{U}_{\text{IP}}(t)=\exp(-i\hat{H}_{\text{IP}}^{S}t)$.
Taking advantage of the Mangnus formula, we can get $\hat{U}_{\text{IP}}(\tau)\simeq\exp(-i\lambda_{\text{eff}}^{2}\hat{J}_{x}^{2}\tau/\Delta_{m})$,
with $\tau=2\pi n/\Delta_{m}$ for the integer number $n$.
This result means that the mechanical mode is decoupled from the NV spins at that moment.
Note that as this operator has no contribution from the mechanical modes, thus
in this instance the system gets insensitive to the states of the mechanical modes.
If the system starts from the initial state $|\psi_{1}^{S}(0)\rangle=|0\rangle_{\text{ph}}|gg......gg\rangle$,
we can obtain the target entangled state for the $N$ NV spins with the form
$|\psi_{\text{T1}}^{\text{NV}}\rangle=[e^{-i\pi/4}|gg......gg\rangle+e^{i\pi/4}|dd......dd\rangle]/\sqrt{2}$, which is the well-known
GHZ state.

We assume these  NV centers are homogeneous and
set $\delta_{dg}^{j}=0$ and $\lambda_{\text{eff}}^{j}=\lambda_{\text{eff}}$ for simplicity.
Then, we can obtain the Hamiltonian in the interaction picture with the form
$\hat{H}_{\text{IP}}^{S}=\lambda_{\text{eff}}(\hat{a}^{\dag}e^{i\Delta_{m}t}+\hat{a}e^{-i\Delta_{m}t})\hat{J}_{x}.$
The dynamics of the system is governed by the unitary evolution operator $\hat{U}_{\text{IP}}(t)=\exp(-i\hat{H}_{\text{IP}}^{S}t)$.
Taking advantage of the Mangnus formula, we can get $\hat{U}_{\text{IP}}(\tau)\simeq\exp(-i\lambda_{\text{eff}}^{2}\hat{J}_{x}^{2}\tau/\Delta_{m})$
when $\tau=2\pi /\Delta_{m}$. This means that the mechanical mode is decoupled from the NV spins at this moment. If the initial  state of the NV centers is $|gg......gg\rangle$,
we can obtain the target entangled state for the $N$ NV spins with the form
$|\psi^{\text{NV}}\rangle=[e^{-i\pi/4}|gg......gg\rangle+e^{i\pi/4}|dd......dd\rangle]/\sqrt{2}$, which is the well-known
GHZ state. The quality of the produced entangled states can be improved significantly by mechanical amplification.

Taking the effective dissipation $\Gamma_{m}^{S}$ and the dephasing
rate $\gamma_{\text{NV}}^{j}$ into consideration, we have the master equation as follow:
\begin{eqnarray}\label{SME30}
\dot{\hat{\rho}}=i[\hat{\rho},\hat{H}_{\text{IP}}^{S}(t)]+\Gamma_{m}^{S}\mathcal{D}[\hat{a}]\hat{\rho}+\sum_{j=1}^N\gamma_{\text{NV}}^{j}\mathcal{D}[\hat{\sigma}_{z}^{j}]\hat{\rho}.
\end{eqnarray}
Then we can make numerical simulations on the dynamical process (to entangle the NV spins) according to equation (\ref{SME30}).
\begin{figure}\label{figs9}
\includegraphics[width=12cm]{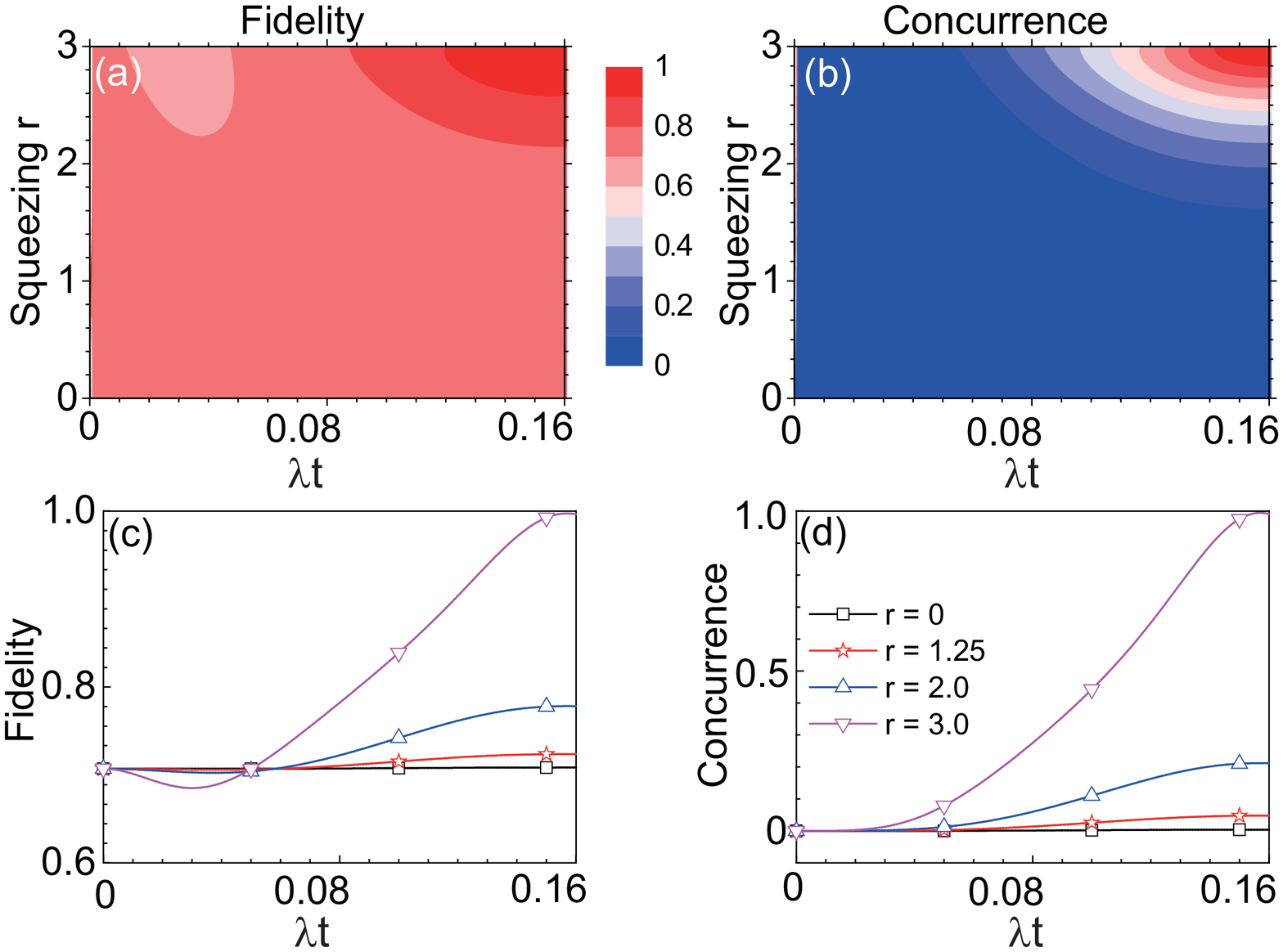}
\caption{\label{fig:wide}(Color online) (a) and (b) Dynamical evolution of the fidelity $F$ and the concurrence $C$ versus the squeezing parameter $r$ for the target entangled state, with the initial state of the NV spins $|gg\rangle$ and the coefficients $\delta_{dg}=0$, $\Delta_{m}=40\lambda$, and $\kappa_{m}^{S}=\gamma_{NV}=0.01\lambda$. (c) and (d) Dynamical evolution of the fidelity $F$
and the concurrence $C$ under different values for the squeezing parameter $r$. }
\end{figure}
Fig. S9  displays the fidelity $F$ of the target entangled states
and the concurrence $C $ for the case of two NV spins varying with the evolution time and the squeezing parameter $r$.
Starting from the initial state $|gg\rangle$,  we can obtain the target entangled state of two NV spins with the form $|\psi_{\text{T1}}^{\text{NV}}\rangle=1/\sqrt{2}[ e^{-i\pi/4}|gg\rangle+e^{i\pi/4}|dd\rangle]$. For a fixed interaction time and in the presence of mechanical dissipation and spin dephasing, we find that, the fidelity $F$ and concurrence $C$,  can be improved significantly by increasing $r$
The quality of the produced entangled state and the speed for generating it can be greatly improved.

Another application of this scheme is to engineer these collective NV spins into the spin squeezed state through the one-axis twisting spin-spin interaction. Due to the computing resources, we choose the number of the NV spins as $N\leq 10$ for numerical simulations.
To quantify the spin squeezed state, we use two different spin squeezing parameters to describe this nonclassical spin state.
First, we utilize the spin squeezed parameter $\xi_{S} ^{2}$
to define the squeezing degree
\begin{equation}\label{SME31}
\xi_{S} ^{2}=\frac{4\min (\Delta \hat{J}_{\vec{n}\perp }^{2})}{N},
\end{equation}%
and this definition is first introduced by Kitagawa and Ueda in 1993 \cite{PhysRevA.47.5138}.
In Eq.~(\ref{SME31}) above, $\vec{n}_{\perp }$ refers to an axis perpendicular to the
mean-spin direction, and the term \textquotedblleft min\textquotedblright\ is the minimization over all directions
$\vec{n}_{\perp }$. The first step is to determine the mean-spin direction $\vec{n}_{0}$ by the
expectation values $\langle \hat{J}_{\alpha }\rangle$, with $\alpha \in
\{x,y,z\}$.  We write $\vec{n}_{0}$ with spherical coordinates $%
\vec{n}_{0}=(\sin \theta \cos \phi, \sin \theta \sin \phi, \cos \theta )$,
and this description is equivalent to the coherent spin state $|\theta,\phi \rangle$.
We can get the other two orthogonal bases which are perpendicular to $\vec{n}_{0}$,
\begin{eqnarray}\label{SME32}
\begin{aligned}
\vec{n}_{1} &=(-\sin \phi ,\cos \phi ,0), \\
\vec{n}_{2} &=(\cos \theta \cos \phi ,\cos \theta \sin \phi ,-\sin \theta ).
\end{aligned}
\end{eqnarray}%
Hence, $\vec{n}_{\perp }=\vec{n}_{1}\cos \beta +\vec{n}_{2}\sin \beta$ is the arbitrary direction vector perpendicular to $\vec{n}_{0}$, and we
can find a pair of optimal quadrature operators by tuning $\beta$. Then we get two components of angular momentum,
\begin{eqnarray}\label{SME33}
\begin{aligned}
\hat{J}_{\vec{n}_{1}} &=-\sin \phi \hat{J}_{x}+\cos \phi \hat{J}_{y}, \\
\hat{J}_{\vec{n}_{2}} &=\cos \theta \cos \phi \hat{J}_{x}+\cos \theta \sin \phi
\hat{J}_{y}-\sin \theta \hat{J}_{z}.
\end{aligned}
\end{eqnarray}%
As a result, we acquire the expression of optimal squeezing parameter
\begin{equation}\label{SME34}
\xi_{S} ^{2}=\frac{2}{N}[\langle \hat{J}_{\vec{n}_{1}}^{2}+\hat{J}_{\vec{n}%
_{2}}^{2}\rangle -\sqrt{(\langle \hat{J}_{\vec{n}_{1}}^{2}-\hat{J}_{\vec{n}%
_{2}}^{2}\rangle )^{2}+4\text{Cov}(\hat{J}_{\vec{n}_{1}},\hat{J}_{\vec{n}_{2}})}],
\end{equation}%
where
\begin{equation}
\text{Cov}(\hat{J}_{\vec{n}_{1}},\hat{J}_{\vec{n}_{2}})=\frac{1}{2}\langle \hat{J}_{\vec{n}_{1}}\hat{J}_{\vec{n}_{2}}+\hat{J}_{\vec{n}_{2}}\hat{J}_{\vec{n}_{1}}\rangle \notag.
\end{equation}

On the other hand, the metrological spin squeezing parameter $\xi_{R} ^{2}$, first introduced by Wineland et al \cite{PhysRevA.46.R6797,PhysRevA.50.67}, can also be applied to describe this squeezed state,
with the relevant definition as
\begin{equation}\label{SME35}
\xi_{R} ^{2}=\frac{N(\Delta\hat{J}_{\perp})^{2}}{\langle\hat{J}_{S}\rangle^{2}}.
\end{equation}%
Furthermore, we note that $\xi_{S}^{2}$ is related to the metrological spin squeezing $\xi_{R}^{2}$ via $\text{min}(\xi_{R}^{2})=[\frac{N}{2\langle\hat{J}_{S}\rangle}]^{2}(\xi_{S}^{2})$,
with the spin length $L_{\text{spin}}=\frac{|\langle\hat{J}_{S}\rangle|}{N/2}$.
Since $|\langle\hat{J}_{S}\rangle|\leq N/2$ and the spin length $L_{\text{spin}}\leq1$,
so we can obtain $\xi_{S}^{2}\leq \xi_{R}^{2}$.
In other words, the metrological spin squeezing $\xi_{R}^{2} < 1$ implies
the spin squeezing  $\xi_{S}^{2}< 1$ according to the definition of Kitagawa and Ueda.
However, the inverse is not true: we can't surely get the metrological spin squeezing $\xi_{R}^{2} < 1$
only through the relation $\xi_{S}^{2}< 1$ \cite{groszkowski2020heisenberglimited,RevModPhys.90.035005,PhysRevLett.116.093602,Wang2019}.

In  quantum metrology,
the metrological gain (the gain of phase sensitivity relative to the standard quantum limit) $G_{m}=(\Delta\theta_{\text{SQL}}/\Delta\theta)^{2}$ is also a figure of merit, where the quantum standard limit
$\Delta\theta_{\text{SQL}}=1/\sqrt{N}$ and the phase uncertainty $\Delta\theta=\xi_{R}/\sqrt{N}$ are
also experimentally achieved in different systems. So we can also obtain the relation $G_{m}=1/\xi_{R}^{2}$.

In a word, we can distinguish spin squeezed states or entangled spin states distinctly for multiple NV centers according to $\xi_{S,R} ^{2}<1$, which also equivalently leads to the direct implications for spin ensemble-based quantum metrology applications as $G_{m}>1$.   And the numerical results in the main text show that these collective NV spins can be steered into the spin squeezed state more quickly as we increasing the squeezing parameter $r$.

\subsection{Enhancing the strain-induced coupling between NV centers and nanomechanical resonators}
This proposed method is also applicable for enhancing the strain-induced spin-phonon coupling through crystal strains in a diamond nanomechanical resonator. We can also achieve the spin-phonon interaction with an exponential enhancement through modifying the spring constant of the nanomechanical resonator.

As illustrated in Fig.~S10(a), we consider the setup consisting of NV centers embedded in a doubly clamped diamond nanomechanical resonator, with dimensions $(l,w,h)$. Electrodes are coated on the  lower surface of the diamond nanobeam. For single NV centers, the ground-state energy level is plotted in Fig.~S10(b), without classical driving, and its Hamiltonian is expressed as
\begin{eqnarray}\label{SME36}
\hat{H}_{\text{NV}}=(D+d_{\|}\epsilon_{\|})\hat{S}_{z}^{2}+g_{e}\mu_{B}\hat{S}_{z}B_{z}
-d_{\bot}[\epsilon_{x}(\hat{S}_{x}\hat{S}_{y}+\hat{S}_{y}\hat{S}_{x})
+\epsilon_{y}(\hat{S}_{x}^{2}-\hat{S}_{y}^{2})],
\end{eqnarray}
where $d_{\|}$ and $d_{\bot}$ are the strain susceptibility parameters parallel and perpendicular to the NV symmetry axis, $\epsilon_{\|}=\epsilon_{z}$, $\epsilon_{\bot}=\sqrt{\epsilon_{x}^{2}+\epsilon_{y}^{2}}$, and $\{\epsilon_{i}\}_{i=x,y,z}$ are the diagonal components of the stain tensor.

Vibration of the diamond nanoresonator periodically changes the local strain at the NV spin's position. This results in a strain-induced electric field, which will act on the corresponding NV center. Here, we focus on the resonant or near-resonant transitions between the states $|\pm1\rangle$ caused by this strain-induced mechanical mode.
Through defining $\hat{\sigma}_{\pm}^{j}=|\pm1\rangle_{j}\langle\mp1|$ and $\hat{\sigma}_{z}^{j}=|+1\rangle_{j}\langle+1|-|-1\rangle_{j}\langle-1|$ for the $j$th NV spin, we can get the $j$th NV spin's Hamiltonian in this two level subspace $\{|+1\rangle,|-1\rangle\}$ with the expression
\begin{eqnarray}\label{SME37}
\hat{H}_{\text{NV}}^{j}=\omega_{m}\hat{a}^{\dag}\hat{a}+\frac{\delta_{B}^{j}}{2}\hat{\sigma}_{z}^{j}
+\lambda^{j}(\hat{a}^{\dag}\hat{\sigma}_{-}^{j}+\hat{a}\hat{\sigma}_{+}^{j}),
\end{eqnarray}
with $\omega_{m}$ the fundamental frequency of this resonator, $\delta_{B}^{j}$ the Zeeman splitting, and the coupling strength $\lambda^{j}/2\pi\sim180~\text{GHz}\times2d_{j}\sqrt{\hbar/ (l^{3}w\sqrt{E\varrho})}/h$.
For simplicity,  here we assume that all of the NV centers are planted near the surface of the diamond resonator with the same distance $d_{j}\simeq 0.5h$.
\begin{figure}
\includegraphics[width=15cm]{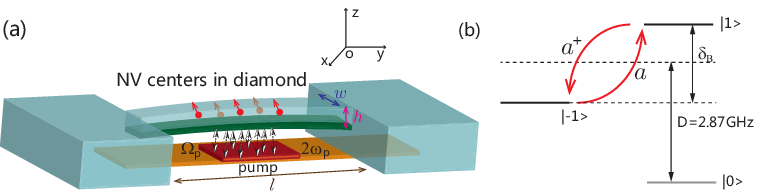}
\caption{\label{fig:wide}(Color online) (a) The strain-induced coupling scheme with NV centers and a doubly clamped diamond nanoresonator. The spring constant of this diamond nanomechanical resonator is capacitively tunable. (b) The ground-state energy-level diagram for single NV center with the strain-mediated phonon-spin transition process($|1\rangle\leftrightarrow|-1\rangle$).  }
\end{figure}

As shown in Fig.~S10(a), the electrode materials are homogeneously coated on the lower surface of the nanobeam,
 and another electrode with a tunable and time-varying voltage is placed just near the lower surface.
The Hamiltonian of this mechanical system with the time-dependent spring constant is expressed as
\begin{eqnarray}\label{SME38}
\hat{H}_{\text{mec}}&=&\frac{\hat{p}_{z}^{2}}{2M}+\frac{1}{2}k(t)\hat{z}^{2}=\frac{\hat{p}_{z}^{2}}{2M}+\frac{1}{2}k_{0}\hat{z}^{2}+\frac{1}{2}k_{r}(t)\hat{z}^{2}.
\end{eqnarray}
The gradient of the electrostatic force from
the electrode has the effect of modifying the spring constant according to $k(t)=k_{0}+k_{r}(t)$,
with $k_{0}=\omega_{m}^{2}M$ the unperturbed fundamental spring constant and the time-varying pump item
$k_{r}(t)\equiv \partial^{2} (C_r V^{2})/(2\partial \hat{z}^{2})=\partial F_{e}/\partial \hat{z}=\Delta k \cos(2\omega_{p}t)$.
Here, $F_{e}=\partial (C_r V^{2})/(2\partial \hat{z})$ is the tunable electrostatic force exerted on the nanobeam by
the electrode, $\hat{z}$ is the displacement, $\Delta k$ is the drive amplitude, and $2\omega_{p}$ is the driving frequency.
The tunable parameters $C_r=\varepsilon_{0}\varepsilon_{r} S/(d+\hat{z})$ and $V=V_{0}+V_{p}\cos2\omega_{p}t$ correspond to the electrode-nanobeam capacitance and
the tunable voltage. Therefore, we can achieve
\begin{eqnarray}\label{SME39}
k_{r}(t)\simeq[\frac{2V_{0}V_{p}\varepsilon S}{d^{2}}]\times\cos2\omega_{p}t=\Delta k\times\cos2\omega_{p}t.
\end{eqnarray}
Defining the displacement operator $\hat{z}=z_{\text{zpf}}(\hat{a}^{\dag}+\hat{a})$
with the zero field fluctuation $z_{\text{zpf}}=\sqrt{\hbar/2M\omega_{m}}$,
we can quantize the Hamiltonian $\hat{H}_{\text{mec}}$ ($\hbar =1$),
\begin{eqnarray}\label{SME40}
\hat{H}_{\text{mec}}&=&\omega_{m}\hat{a}^{\dag}\hat{a}- \Omega_{p}\cos(2\omega_{p}t)(\hat{a}^{\dag}+\hat{a})^{2},
\end{eqnarray}
where $\omega_{m}=\sqrt{k_{0}/M}$ is the fundamental frequency, and $\Omega_{p}=-\Delta k z_{\text{zpf}}^{2}/2=2V_{0}V_{p}\varepsilon S/(d+\hat{z})^{2} \times z_{\text{zpf}}^{2}/2$ is the nonlinear drive amplitude.
As a result, utilizing this method, we can obtain the second-order nonlinear interaction through modulating the spring constant in time.

In this case,  we can obtain the total Hamiltonian with the same expression as the magnetic coupling scheme
\begin{eqnarray}\label{SME41}
\hat{H}_{\text{Total}}&\simeq& \delta_{m}\hat{a}^{\dag}\hat{a}-\frac{\Omega_{p}}{2}(\hat{a}^{\dag 2}+\hat{a}^{2})+\sum_{j=1}^{N}[\frac{\delta_{\pm}^{j}}{2}\hat{\sigma}_{z}^{j}
+\lambda(\hat{a}^{\dag}\hat{\sigma}_{-}^{j}+\hat{a}\hat{\sigma}_{+}^{j})],
\end{eqnarray}
where the coefficients are respectively $\delta_{m}=\omega_{m}-\omega_{p}$ and $\delta_{\pm}^{j}=\delta_{B}^{j}-\omega_{p}$.
Considering the Hamiltonian (\ref{SME40}),
we can also diagonalize the mechanical mode by the unitary transformation $\hat{U}_{s}(r)=\exp[r(\hat{a}^{2}-\hat{a}^{\dagger2})/2]$.
Define the squeezing parameter $r$ via the relation $\tanh2r=\Omega_{p}/\delta_{m}$.
As a result, we obtain the Rabi Hamiltonian effectively in the squeezed frame,
\begin{eqnarray}\label{SME42}
\hat{H}_{\text{Rabi}}^{N}=\Delta_{m}\hat{a}^{\dag}\hat{a}+\sum_{j=1}^{N}[\frac{\delta_{\pm}^{j}}{2}\hat{\sigma}_{z}^{j}
+\lambda_{\text{eff}}(\hat{a}^{\dag}+\hat{a})\hat{\sigma}_{x}^{j}].
\end{eqnarray}
The coefficients $\Delta_{m}=\delta_{m}/\cosh 2r$ and $\delta_{\pm}^{j}$ correspond to the free Hamiltonian of the mechanical mode and the NV spins in the squeezed frame.
Furthermore, we can also obtain the exponentially enhanced spin-phonon coupling strength
$\lambda_{\text{eff}}\sim\lambda e^{r}$ in this new frame,
which can be comparable with $\Delta_{m}$ and $\delta_{\pm}^{j}$, even stronger than both of them.
As discussed in the previous section, we can easily tune the amplitude $\Omega_{p}$ of this nonlinear pump through modifying $\varepsilon_{r}$, $V_{0}V_{p}$, $S$, and $d$.
Therefore in this scheme we can achieve $\Omega_{p}$ varying with the regime from $\sim 2\pi\times1$kHz to $\sim 2\pi\times0.1$GHz. As a result, we can explore the
same idea to enhance the spin-phonon and spin-spin interactions in this strain coupling system.

%

\end{document}